\documentclass[a4paper,twocolumn,11pt,accepted=2026-04-10]{quantumarticle}

\pdfoutput=1

\usepackage[utf8]{inputenc}
\usepackage[T1]{fontenc}
\usepackage[english]{babel}
\usepackage{url}

\usepackage{textcomp}
\usepackage{amsmath,amssymb,mathtools,mathrsfs}
\usepackage{physics}
\usepackage{multirow}
\usepackage{booktabs,tabularx,makecell}
\usepackage{placeins}
\usepackage{units}
\usepackage{tikz}
\usepackage[normalem]{ulem}
\usepackage{xcolor}

\definecolor{girlishpink}{RGB}{211,129,195}

\usepackage[
  colorlinks=true,
  linkcolor=teal,
  citecolor=blue,
  urlcolor=girlishpink
]{hyperref}
\usepackage[capitalise,noabbrev,nameinlink]{cleveref}

\begin{document}
	\title{
 Comparing the performance of practical two-qubit gates for individual $^{171}$\text{Yb}
 ions in yttrium orthovanadate}
	
\author{Mahsa Karimi}
\email{mahsa.karimi1@ucalgary.ca}
\affiliation{Institute for Quantum Science and Technology, and Department of Physics \& Astronomy, University of Calgary, 2500 University Drive NW, Calgary, Alberta T2N 1N4, Canada}

\author{Faezeh Kimiaee Asadi}
\affiliation{Institute for Quantum Science and Technology, and Department of Physics \& Astronomy, University of Calgary, 2500 University Drive NW, Calgary, Alberta T2N 1N4, Canada}

\author{Stephen C. Wein
}\thanks{Current affiliation is Quandela, 7 Léonard De Vinci, 91300 Massy, France.}

\affiliation{Institute for Quantum Science and Technology, and Department of Physics \& Astronomy, University of Calgary, 2500 University Drive NW, Calgary, Alberta T2N 1N4, Canada}

\author{Christoph Simon}
\affiliation{Institute for Quantum Science and Technology, and Department of Physics \& Astronomy, University of Calgary, 2500 University Drive NW, Calgary, Alberta T2N 1N4, Canada}

\begin{abstract}
In this paper, we investigate three schemes for implementing Controlled-Z (CZ) gates between individual ytterbium (Yb) rare-earth ions doped into yttrium orthovanadate (YVO$_4$ or YVO). Specifically, we investigate the CZ gates based on magnetic dipolar interactions between Yb ions, photon scattering off a cavity, and a photon interference-based protocol, with and without an optical cavity.
We introduce a theoretical framework for precise computations of state and gate infidelities, accounting for noise effects.
We then compute the state fidelity for each scheme to evaluate the feasibility of their experimental implementation. Based on these results, we compare the performance of the two-qubit gate schemes and discuss their respective advantages and disadvantages. 
We conclude that the probabilistic photon interference-based scheme offers the best fidelity scaling with cooperativity and is superior with the current technology of Yb values, while photon scattering is nearly deterministic but slower with less favourable fidelity scaling as a function of cooperativity. The cavityless magnetic dipolar scheme provides a fast, deterministic gate with decent fidelities if close ion localization can be realized. While focusing on $^{171}\text{Yb}^{3+}$: YVO system as a case study, the theoretical tools and approaches developed in this work are broadly applicable to other spin qubit systems.
\end{abstract}

\maketitle

\section{Introduction}

Two-qubit gates are fundamental for enabling entanglement swapping in quantum repeaters \cite{jennewein2001experimental, sangouard2011quantum} and are crucial for generating and mapping entangled states in distributed quantum computing \cite{deutsch1989quantum}. 
From a computational standpoint, these gates are essential for universal quantum computation, as they enable the efficient decomposition of arbitrary quantum gates into circuits composed of Controlled-Z (CZ) and single-qubit gates \cite{nielsen2001quantum, kitaev2002classical}.

For the practical implementation of quantum computing, various experimental platforms have been explored, including solid-state systems such as superconducting qubits \cite{devoret2013superconducting}, quantum dots \cite{loss1998quantum}, impurity-doped solids like  nitrogen-vacancy (NV) centers in diamonds \cite{wrachtrup2006processing}, silicon-vacancy (SiV) centers \cite{sukachev2017silicon}, T-centers \cite{simmons2024scalable}, and rare-earth ions (REIs) embedded in crystals \cite{kinos2021roadmap,longdell2005stopped, zhong2015optically}. Superconducting qubits offer fast gate speeds and easy integration with microwave circuits, but suffer from short coherence times, fabrication variability, and cryogenic requirements.
Quantum dots feature strong optical transitions but broad, unstable linewidths and short spin coherence, requiring cryogenics and cavities.
NV centers offer long spin coherence and strong optical transitions but suffer from phonon sidebands and linewidth broadening at higher temperatures. SiV centers have reduced phonon sidebands but need ultra-low temperatures for good spin coherence. T-centers provide narrow transitions, long coherence, and telecom emission, yet are prone to strain- and charge-induced spectral diffusion. Rare-earth ions exhibit weak but narrow optical transitions, long spin coherence, and strong memory potential with minimal spectral diffusion, though they require cryogenic operation.
The relatively low radiative decay rate of REIs \cite{dibos2018atomic} can be significantly improved by coupling them to a nanophotonic optical cavity, which enhances their emission through Purcell enhancement \cite{zhong2015nanophotonic, ourari2023indistinguishable}. Additionally, incorporating a cavity boosts photon collection efficiency and improves single-photon indistinguishability.

Within rare-earth ions (REIs), Kramers ions, which have an unpaired electron in their 
$f$ shell such as erbium (Er$^{3+}$), and ytterbium (Yb$^{3+}$) exhibit substantial magnetic moments on the order of the Bohr magneton $(\mu_B)$. Thus, these ions can interact with nearby ions through magnetic dipole-dipole interactions. 
Er$^{3+}$ is known for telecom-compatible networking due to its native emission wavelength, while Yb$^{3+}$ offers simpler level structure, comprising of two electronic multiplets for the ground ($^{2}$F$_{7/2}$) and excited ($^{2}$F$_{5/2}$) states. Notably, $^{171}\text{Yb}^{3+}$ isotope of Yb$^{3+}$ is the only Kramers rare-earth ion with the lowest nuclear spin of $I = 1/2$. Hence, it benefits from long coherence times of nuclear levels \cite{nicolas2023coherent}, while maintaining the simplest possible hyperfine energy level structure with only a few narrow transitions in the optical absorption spectra \cite{tiranov2018spectroscopic, kindem2018characterization}. This simplicity facilitates more efficient manipulation of spin qubits and gate operations. Consequently, Yb is an attractive candidate for practical applications in rare-earth quantum information processing \cite{welinski2016high, ortu2018simultaneous, 
welinski2020coherence,
businger2020optical, businger2022non, chiossi2024optical,
ranon1968paramagnetic,
krankel2004continuous,
bartholomew2020chip, 
ruskuc2022nuclear, xie2024scalable}.

Among host crystals, Yb ions doped into yttrium orthosilicate (Y$_2$SiO$_5$ or YSO) and yttrium orthovanadate (YVO$_4$ or YVO) have been extensively studied \cite{welinski2016high,tiranov2018spectroscopic,kindem2018characterization}. YSO is widely used in quantum experiments, primarily due to the small magnetic moments of its components \cite{equall1994ultraslow}. Conversely, YVO is of interest due to its high site symmetry, leading to narrow inhomogeneous linewidths when doped with $^{171}\text{Yb}^{3+}
$ \cite{kindem2018characterization}.
YVO is also promising for fabricating nanophotonic cavities \cite{zhong2016high}, and particularly for Yb:YVO 
\cite{kindem2020control,wu2023near}, making it a compelling choice for this study.

Thus far, the design of quantum gates with large ensembles of REIs has been studied \cite{fraval2005dynamic, rippe2008experimental, longdell2004demonstration}. However, scalability issues with these designs \cite{wesenberg2007scalable} have prompted a shift toward using individual ions \cite{walther2009extracting, walther2015high, kinos2021designing, kinos2022high}. 
In that regard, significant advancements have been made in addressing individual rare-earth ions, particularly with praseodymium \cite{kolesov2012optical}, erbium \cite{yin2013optical, yang2023controlling}, ytterbium \cite{kindem2020control}, and neodymium ions coupled to a photonic crystal resonator \cite{zhong2018optically}. Recent experimental progress has achieved probabilistic generation of maximally entangled state between individual $^{171}\text{Yb}^{3+}$ ions located in remote cavities \cite{ruskuc2025multiplexed}. Despite these advances, there has yet to be an experimental implementation of CZ gates between single REIs.

To implement two-qubit gates between individual rare-earth ions, various interaction mechanisms have been proposed. These include electric \cite{ohlsson2002quantum, asadi2018quantum}, magnetic dipolar \cite{grimm2021universal, kinos2021roadmap}, and cavity-mediated \cite{asadi2020cavity} interactions.

In \cite{grimm2021universal}, the authors have compared Controlled-NOT (CNOT) gates based on phase-accumulation and blockade mechanisms utilizing magnetic dipole-dipole interaction. They have shown that leveraging the full interaction strength, the magnetic dipolar phase-based scheme is faster and less constrained by the distance between ions compared to the magnetic dipolar blockade gate scheme.
In contrast to dipolar gates, cavity-mediated interaction gates do not require qubits to be in close proximity. Instead, these gates rely on the presence of a cavity, which may also be necessary for other purposes, such as improving interfaces between stationary qubits (like quantum emitters) and flying qubits (like photons).
Therefore, it is important to explore and compare the advantages of cavity-based and cavityless two-qubit gates between individual Yb ions, evaluating their implementation feasibility more thoroughly.

In this paper, our goal is to guide experimental efforts by comparing gate schemes that hold practical promise in near to medium term, given the current state of the technology for implementation in the Yb:YVO system. Specifically, our study covers different types of approaches—namely, direct interaction, cavity-assisted, and interference-based schemes. These schemes included a broad range of protocols that differ not only in their reliance on cavity versus cavity-free implementations, but also span the range from probabilistic to near-deterministic and deterministic operation. Depending on the specific goals and constraints of an experiment, different schemes may be prioritized. We focus on the impact of intrinsic atomic properties that fundamentally limit gate performance, while noting that technological issues such as collection efficiencies still need to be improved in practice as well.

A key figure of merit for assessing the performance of a gate scheme is the fidelity $(F)$. Notably, $1/(1-F)$ provides an estimate of the number of gate operations that can be performed without the need for quantum error correction \cite{childs2017lecture, grimm2021universal}.  Consequently, high-fidelity gates are essential for fault-tolerant quantum computing and first-generation quantum repeaters.
To understand the sensitivity of each gate scheme to error parameters, we develop a theoretical framework to compute fidelity. We model the evolution of our system using the Gorini–Kossakowski–Sudarshan–Lindblad (GKSL) equations \cite{lindblad1976generators}. By computing perturbed solutions to the corresponding GKSL equations, we derive closed-form expressions for calculating infidelities. 
Our perturbative approach captures noise-induced errors through a formulation that relies merely on the noise parameters and the solution of the noiseless system, thus avoiding the need to solve the GKLS master equation directly. Therefore, our results offer a more feasible computational method compared to existing literature \cite{wein2020analyzing, asadi2020cavity, asadi2020protocols}.
Concretely, denoting the number of levels in our system needed for the gate operation by $d$, our method's computational cost scales as $O(d^3)$ as opposed to $O(d^6)$ being the cost of computation in prior works. Furthermore, as highlighted by \cite{blume2022taxonomy}, the connection between the Lindbladian and error rates is still being uncovered, and our work can be considered one step further in that direction. These
results can be applied to estimate single- and two-qubit gate errors of a given system,
laying the foundation for universal quantum computing.

Using the above-mentioned perturbative approach we report the state and gate fidelies of the magnetic dipolar gate scheme presented in this work. We then discuss the requirements for implementing these gates within the Yb:YVO system, comparing the performance of each scheme based on current experimental technology. We also highlight their respective advantages and disadvantages, providing a comparative framework to guide the selection of the most suitable approach.

We note that while this comparison study focuses on the $^{171}\text{Yb}^{3+}$:YVO system, the mathematical tools and methodologies presented in this work can be applied to other spin qubits, including other rare-earth ions in various host materials and defect centers in materials such as diamond, silicon, or silicon carbide.

The paper is structured as follows:
In \cref{yb system}, we introduce the Yb ion system. \cref{gates} investigates the two-qubit gate schemes. The details of the perturbative approach for computing the fidelities are presented in \cref{sec:state-fid}. In \cref{imp}, we analyze the fidelity calculations for the gate schemes. We also discuss the practical implementation of key parameters influencing the fidelities, along with the current achievable fidelity for Yb ions. In \cref{comparing}, we compare the different gate schemes and provide the pros and cons of each. Finally, we conclude and present an outlook in \cref{disc}.

\section{Yb rare-earth ion properties}\label{yb system}

\begin{figure}[t]
\centering
\includegraphics[width=0.5\textwidth]{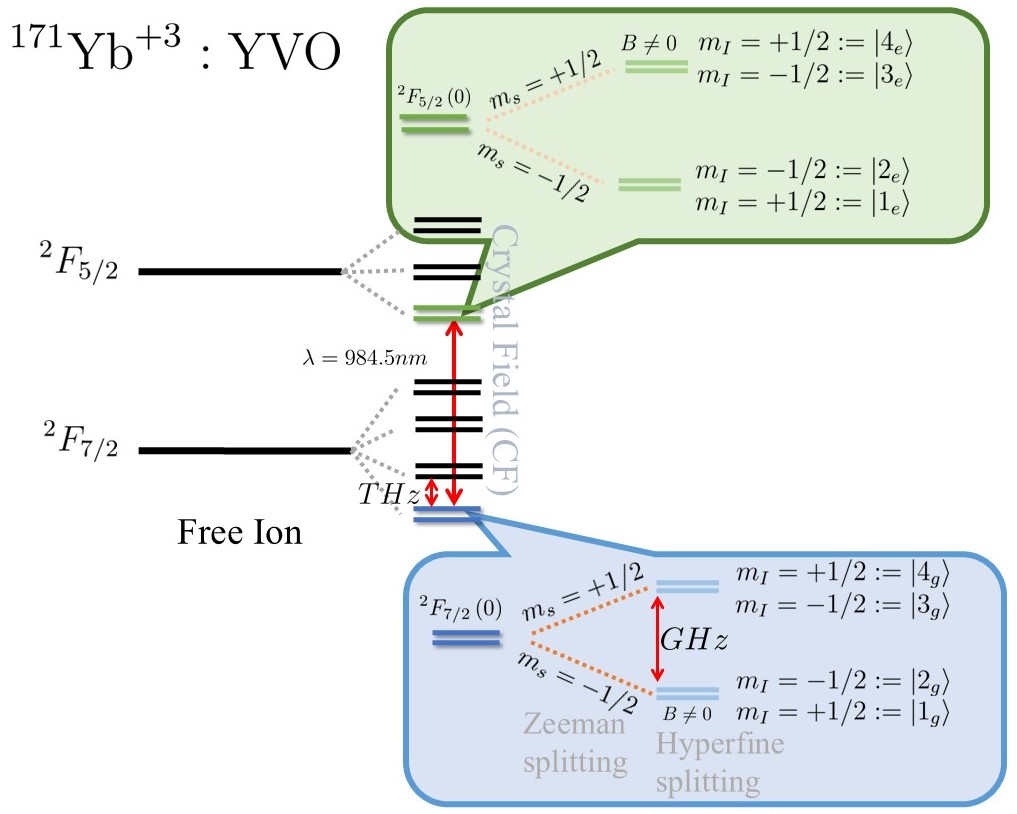}
\caption{The level structure scheme of $^{171}\text{Yb}^{3+}$
ions doped into YVO crystal, where a high magnetic field separates the electron and nuclear spin mixing.}\label{Yb structure}
\end{figure}

$^{171}\text{Yb}^{3+}$:YVO has four Kramers doublets in the ground state and three in the excited state. At low temperatures, only the lowest doublet is populated. \cref{Yb structure} illustrates the energy level structure of this system in the presence of a magnetic field \cite{kindem2018characterization, huan2019characterization}.
Within a Kramers doublet, 
the effective spin Hamiltonian of the Kramers ions with non-zero nuclear spin, $\boldsymbol{I}$, can be written as~\cite{abragam2012electron}

\begin{equation}\label{H_eff}
\begin{split}
 H_\mathrm{eff} = &{\mu _B}\mathbf{B\cdot g\cdot S} - {\mu_N}\mathbf{B\cdot g _N\cdot I}\\
 &+\mathbf{A\cdot S\cdot I}+\mathbf{I\cdot Q\cdot I},
 \end{split}
\end{equation}
where $\mu_B(\mu_N)$ is the electronic (nuclear) Bohr magneton, $\mathbf{B}$ is the external magnetic field, $\mathbf{S(I)}$ is the electronic (nuclear) spin operator, $\mathbf{g}(\mathbf{g_N})$ is the electronic (nuclear) Zeeman tensor, $\mathbf{A}$ is the hyperfine tensor, and $\mathbf{Q}$ is the electronic-quadrupole tensor.
For $^{171}\text{Yb}$, $\mathbf{I\cdot Q\cdot I}=0$, because this term only appears for ions with $I \ge 1$ facilitating the spectroscopic properties analysis of this isotope comparing to $^{167}\text{Er}^{3+}$ with $I=7/2$ \cite{tiranov2018spectroscopic}.

In this paper (similar to the other Yb:YVO quantum interference experiments), we restrict ourselves to the 
lowest crystal-field (CF) levels in the ground and excited states ($^2F_{7/2}(0)\leftrightarrow$ $^2F_{5/2}(0)$) coupled through an optical transition at 
$\sim984.5$ nm. 
The radiative decay from $^2F_{5/2}(0)$ can proceed to all four ground states within the same CF level (labelled $^2F_{7/2}$ in \cref{Yb structure}), with a branching ratio of $\beta = 0.45$ into the lowest ground state level $^2F_{7/2}(0)$. To enhance this transition, the experiment exploits Purcell enhancement by tuning the cavity to resonate exclusively with $^2F_{5/2}(0) \leftrightarrow$ $^2F_{7/2}(0)$ transition, while remaining detuned from all other CF transitions \cite{kindem2020control, ruskuc2024single}.

We note that superhyperfine interactions, specifically the magnetic dipole-dipole interaction between Yb and host nuclei may cause fluctuation in the optical transitions. 
However, for this system, advanced techniques such as dynamical decoupling sequences are employed to suppress interactions with the nuclear spin bath and extend the coherence time \cite{kindem2020control}. Therefore, we focus on the simplified hyperfine energy level structure of this system, shown in \cref{Yb structure}. Although we do not consider the interaction between Yb ions and the spin bath here, we note that the collective state of the surrounding Vanadium nuclear spins $^{51}\mathrm{V}$, with their weak coupling to the environment, offers a promising resource for long-lived quantum memory applications \cite{ruskuc2022nuclear}.

\section{gate scheme descriptions}\label{gates}

In this section, we discuss practical two-qubit gate
schemes involving individual $^{171}\text{Yb}^{3+}$ ions doped into YVO crystals (see \cref{table-gates}).
With a focus on near- to medium-term feasibility based on the current state of technology, we outline the rationale behind our scheme selection.

To provide a general roadmap suitable for different parts of a quantum network depending on specific requirements, we pick schemes that are good representatives for different potential approaches by including a direct interaction scheme, a cavity-mediated scheme, and an interference-based scheme. Among deterministic two-qubit gates, those based on the dipolar interactions have the advantage of not requiring a cavity for their
implementation. However, for Yb:YVO, there is no information on the difference in the permanent electric dipole moment between the ground and excited states ($\Delta\mu$), mainly due to the lack of a first-order DC Stark shift, a limitation arising from the site symmetry \cite{kindem2020control}. This inherent characteristic significantly hinders the implementation of electric dipolar gates, including both blockade-based and phase-accumulation schemes, in the Yb:YVO system.
Nevertheless, magnetic dipolar interaction-based gates, such as blockade and phase-accumulation schemes, remain viable for this system. However, as shown in \cite{grimm2021universal}, a comparison of these two approaches reveals that the magnetic blockade gate, which relies on frequency shifts, is a promising scheme for ions' separation in the order of $1$ nm. Given the challenge of identifying such closely spaced Yb ion pairs, we focus on the magnetic phase-accumulation scheme, which provides relatively greater flexibility in terms of separation distance between the ions and operates significantly faster than the magnetic blockade gate \cite{grimm2021universal}.

Among cavity-mediated two-qubit gates, as demonstrated in \cite{asadi2020cavity}, the photon scattering achieves higher fidelity with lower cavity cooperativity requirements. This is particularly relevant for the Yb:YVO system (and more generally for REIs), where achieving high cooperativity is challenging. Although the photon scattering gate scheme is intrinsically deterministic, a non-ideal photon detection process may result in a gate efficiency of less than one \cite{duan2005robust}. 
Among probabilistic schemes, both the single-photon heralding protocol \cite{cabrillo1999creation} and the two-photon entanglement protocol \cite{barrett2005efficient} have been implemented in the Yb:YVO system \cite{ruskuc2025multiplexed}. The single-photon heralding scheme (also called photon-number encoding or the \textsc{N} protocol) scales linearly with photon loss probability $\eta$, whereas the two-photon entanglement protocol (time-bin encoding or Barrett–Kok protocol) scales quadratically with $\eta$. Despite the improved efficiency scaling of the \textsc{N} protocol, its lack of loss detection reduces the fidelity of Bell pairs. This can be countered by decreasing the excitation power, albeit with a reduction in efficiency.
Experimentally, Ref. \cite{ruskuc2025multiplexed} demonstrated a fidelity–rate trade-off between the two schemes: the single-photon heralding protocol achieved F$=0.72$ with a rate of R$=3.1$ Hz, while the two-photon scheme reached F$=0.81$ at a much lower rate of R$=0.049$ Hz. We note that the improvement in the collection and detection efficiencies can subsequently improve the low rates. Furthermore, the two-photon protocol is inherently insensitive to unknown optical phase noise, whereas the single-photon scheme requires active phase stabilization.
For these reasons, in this work, we focus on the two-photon entanglement heralding protocol.
The other entanglement generation schemes, such as those discussed in \cite{beukers2024remote}, may be interesting to explore in future work in order to better assess their practical feasibility and potential implementation challenges. 

Herein, following Ref. \cite{grimm2021universal}, we consider the magnetic dipolar interaction of Yb ions to implement a two-qubit gate between two ground states by inducing a phase difference between two excited states within the ions.
Next, following Ref. \cite{asadi2020cavity}, we discuss a cavity-mediated interaction, including scattering a single photon off a cavity-ion system to execute a CZ gate between two Yb ions.
Finally, based on \cite{barrett2005efficient}, we explore a photon interference-based entanglement generation protocol that can be adapted into a CZ gate between Yb ions. We demonstrate that incorporating a cavity can enhance the fidelity of this gate.

\begin{table*}[t]
\small
\centering
\resizebox{\textwidth}{!}{%
\begin{tabular}{||c|c|c|c|c|c||}
\hline
Gate Schemes & Determinism  & Medium & Dominant error & Infidelity scaling & \# Levels\\
\hline
  Magnetic dipolar  & deterministic   & crystal bulk  & ions' separation & $O(r^3)$ & 4 \\
\hline
  Photon scattering & near deterministic  & cavity & cooperativity & $O(C^{-\frac{2}{3}})$ & 3\\
\hline
  Photon interference-based  &  probabilistic  & bulk or cavity & cooperativity & $O(C^{-1})$ &3 \\
\hline
\end{tabular}%
}
\caption{Two-qubit gate schemes discussed in this paper based on their requirements to be performed. The dominant sources of error, measured via fidelity, are written for each gate. Furthermore, we provide the infidelity scaling for the most dominant error factor. For the magnetic dipolar gate, the infidelity scaling $O(r^3)$ is valid if the direct interaction during activation can be ignored (see \cref{imp} for more details). A comparison between achievable fidelities and corresponding gate times of these gates is shown in \cref{fig:gate time}.}
\label{table-gates}
\end{table*}

\subsection{Magnetic dipolar interaction gate scheme}\label{sec:magnetic-dipolar}

\begin{figure}[t]
\centering
\includegraphics[width=0.5\textwidth]{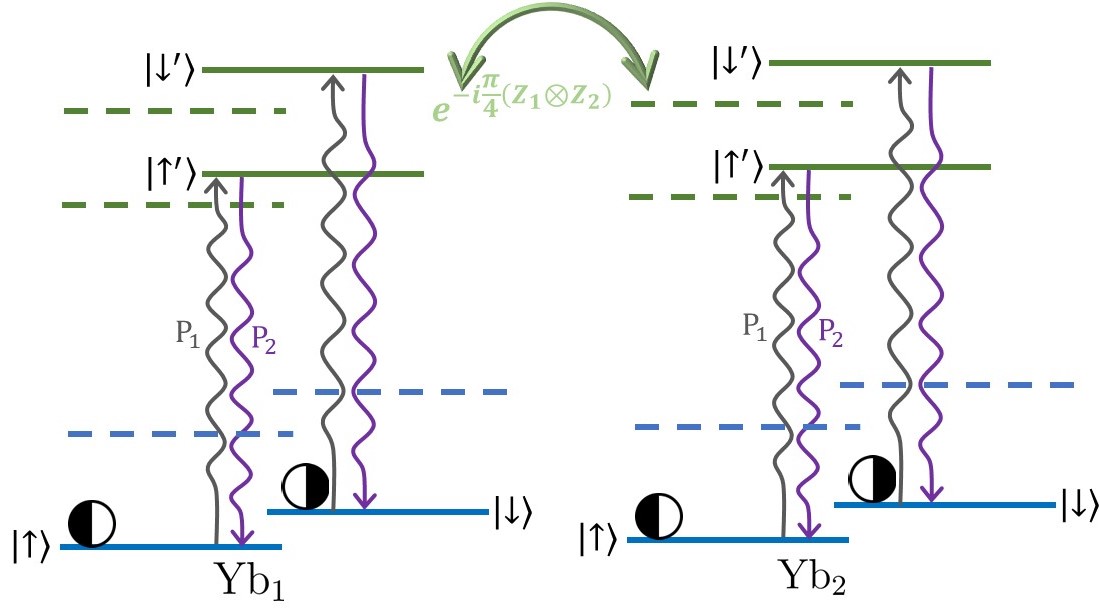}
\caption{Diagram of the magnetic dipolar gate mechanism. Here the passive and active qubit levels (i.e., $\ket{\uparrow}, \ket{\downarrow}$ and $\ket{\uparrow'}, \ket{\downarrow'}$) are defined with hyperfine levels $\ket{1_g}, \ket{2_g}$ and $\ket{2_e}, \ket{4_e}$ presented in \cref{Yb structure}. The pulse sequence for executing a magnetic dipolar-based phase gate between nearby Yb ions begins with preparing a superposition of passive qubit states indicated by half-black coloured circles. The population is then transferred to the excited state using four simultaneous $\pi$ pulses, denoted by $P_1$.  A two-qubit phase gate is applied through electronic magnetic dipolar interaction. After a time delay, the population is returned to the ground state using another set of $\pi$ pulses, denoted by $P_2$.} 
\label{yb-qubits}
\end{figure}

As suggested in \cite{grimm2021universal,kinos2021roadmap}, one can exploit the magnetic dipolar interaction between two Yb ions to perform a two-qubit phase gate.
In our setup, we consider two nearby Yb ions doped into YVO. For each ion, we define a ``passive qubit" by utilizing two of the lowest hyperfine ground state energy levels that share the same electron spin but differ in nuclear spins. We denote these levels as $\ket{\uparrow}$ and $\ket{\downarrow}$, as depicted in \cref{yb-qubits}. 
These passive qubits have a long coherence time, and therefore can serve as memory qubits. However, for the gate operation, 
to speed up the slow microwave couplings, mitigate microwave-induced interactions on nearby ions and enhance the interaction strength,
the quantum state of passive qubits must be transferred to ``active qubits" in the excited state, which possess a significantly different g-tensor compared to the ground state \cite{grimm2021universal,kinos2021roadmap}.
For each Yb, we define active qubit levels using a different electronic spin level. This selection allows the active qubits to interact via the electronic magnetic dipolar interaction, which is proportional to $\mu_B$, rather than the nuclear magnetic dipolar interaction, which is proportional to $\mu_N$ \cite{grimm2021universal}.  Here, we define the active qubit levels using solid green lines and denote them by $\ket{\uparrow'}$ and $\ket{\downarrow'}$ as shown in \cref{yb-qubits}. We describe this selection in \cref{impl:MD}.
In the following, we examine the magnetic dipolar interactions between active qubits to design a two-qubit gate between two Yb ions.

After initializing each Yb ion into the lowest hyperfine ground state ($\ket{\uparrow}$), and applying a $\pi/2$ microwave pulse to create a superposition of the $\ket{\uparrow}$ and $\ket{\downarrow}$ states, the two-qubit gate process begins.
First, the population is transferred from the passive qubits to the active qubits by applying four $\pi$ pulses ($P_1$) simultaneously (two on each ion) as shown in \cref{yb-qubits}. This results in the transitions $\ket{\uparrow} \mapsto \ket{\uparrow'}$ and $\ket{\downarrow} \mapsto \ket{\downarrow'}$, placing each Yb ion into a superposition of $\ket{\uparrow'}$ and $\ket{\downarrow'}$. At this stage, the magnetic dipole-dipole interaction 
between the active qubits performs a two-qubit phase gate. If the first qubit is in the state $\ket{\uparrow'}$, the dipole-dipole interaction induces $\exp(i\frac{\pi}{4} Z_2)$ on the second qubit, and if the first qubit is in the state $\ket{\downarrow'}$ it induces $\exp(-i\frac\pi4 Z_2)$ on the second qubit. Finally, after a time delay, the quantum states are brought back to the passive qubits by applying another set of $\pi$ pulses ($P_2$). A schematic representation of this process is shown in \cref{yb-qubits}.

Note that to perform a CZ gate between passive qubits, additional single-qubit gates are required in conjunction with the above phase gate

\begin{align}
\mathrm{CZ}^{(1)} = (\sqrt{Z}\otimes\sqrt{Z}) U_1
\end{align}
where $\sqrt Z$ for a Pauli $Z$ is defined as $\sqrt{Z}:= \exp(-i\frac{\pi}{4} Z)$, and $U_{1} = \exp( i\frac{\pi }{4}(Z\otimes Z))$ is the unitary implemented by the Ising interaction. Once we have a CZ gate, a CNOT gate also can be achieved by applying two Hadamard gates on the target qubit $\mathrm{CNOT}=(\mathbb I\otimes H) \mathrm{CZ} (\mathbb I\otimes H)$.
Single qubit gates can be performed on passive qubits either through microwave fields or with the help of active qubits through optical fields \cite{grimm2021universal}.\\

\subsection{Photon scattering gate scheme}\label{Duan-Kimble}

Scattering a single photon from the qubit-cavity system can be used to perform a controlled phase-flip gate between qubits \cite{duan2005robust,lin2006one}. This method has been experimentally realized as a locally controlled phase-flip gate with neutral atoms \cite{welte2018photon} and has been theoretically explored for non-local controlled phase gates between rare-earth ions in separate microsphere cavities \cite{xiao2004realizing}. While these studies focus on the strong coupling regime, Ref \cite{asadi2020cavity} extends the approach to implement a CZ gate between two qubits and provides a fidelity expression applicable to both weak and strong coupling regimes.

This scheme requires the detection of a single photon. Alternatively, it can be adapted for quantum non-demolition (QND) measurements to detect the presence of the photon without measuring or destroying its quantum state by probing the quantum emitter system \cite{o2016nondestructive}.\\

\begin{figure}[t]
\centering
\includegraphics[width=0.5\textwidth]{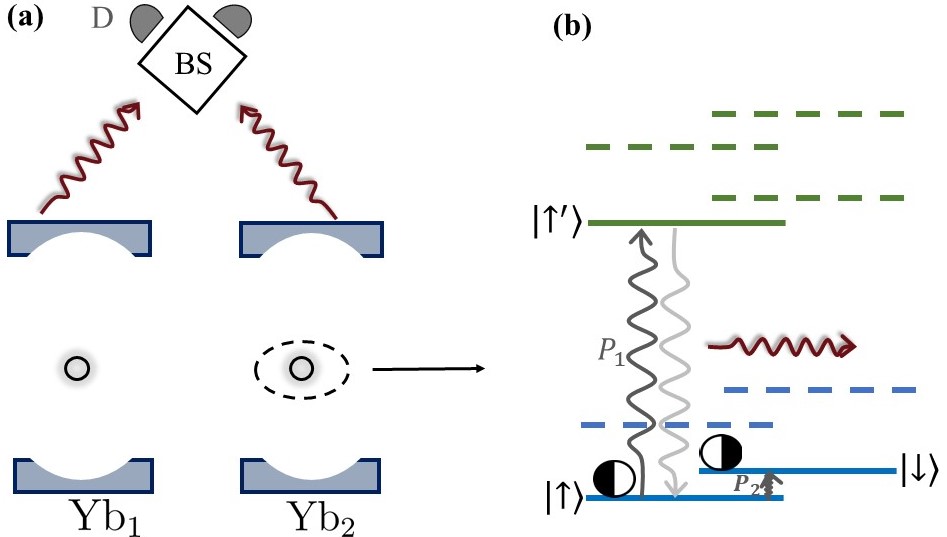}
\caption{(a) Schematic of the photon-scattering gate scheme setup. (b) The pertinent hyperfine energy level structure of the Yb ions taken from \cref{Yb structure}, and the coupling configuration, where the $\ket{\uparrow}\mapsto\ket{\uparrow'}$ transitions of both Yb ions are resonant with each other and with the cavity mode $\omega_c$.} 
\label{Duan-Kimble scheme}
\end{figure}

For this scheme, we employ two of the ground state hyperfine levels of Yb ions as qubit levels ($\ket{\uparrow}$ and $\ket{\downarrow}$), while the first hyperfine level in the excited state serves as an ancillary level ($\ket{\uparrow'}$). The interaction between two Yb ions is mediated by scattering a single photon (in the state $\ket{P}$) from a single-sided cavity containing the Yb ions, as illustrated in \cref{Duan-Kimble scheme}. We tune the $\ket{\uparrow} \mapsto \ket{\uparrow'}$ transition of the ions to resonance with each other and with the cavity. 
If both qubits are in the state $\ket{\downarrow}$, the photon enters the cavity, reflects off its interior, and exits, resulting in a $\pi$-phase shift in the joint ion-photon state. Conversely, if either or both qubits are in the state $\ket{\uparrow}$, the cavity mode is altered, preventing the photon from entering, and the photon is instead reflected off the cavity's out-coupling mirror. Therefore, a photon detector (denoted by D in \cref{Duan-Kimble scheme}) can herald the CZ gate. The unitary operator describing this scheme is given with
$\mathrm{CZ^{(2)}} = {\exp({i\pi | { \downarrow  \downarrow }\rangle \left\langle { \downarrow  \downarrow } \right| \otimes \left| P \right\rangle \left\langle P \right|})}$. A CNOT gate is achieved by post-selecting the state of the photon and conjugated by Hadamard gates on the target qubit.

\subsection{Photon interference-based gate scheme}

This scheme, originally introduced by Barrett and Kok in \cite{barrett2005efficient}, is a non-deterministic entanglement generation protocol that can be applied in designing quantum network elements. \cite{reiserer2016robust, asadi2018quantum,
asadi2020protocols, ji2022proposal}. While initially developed without a cavity, applying remote cavities to host each individual ion can significantly enhance the fidelity of the scheme (see \cref{implementation:BK scheme} for more details). This entanglement generation protocol can be adapted into a CZ gate by measuring the photons in the mutually unbiased basis (MUB) instead of the photon number basis \cite{lim2005repeat}.

\begin{figure}[t]
\centering
\includegraphics[width=0.5\textwidth]{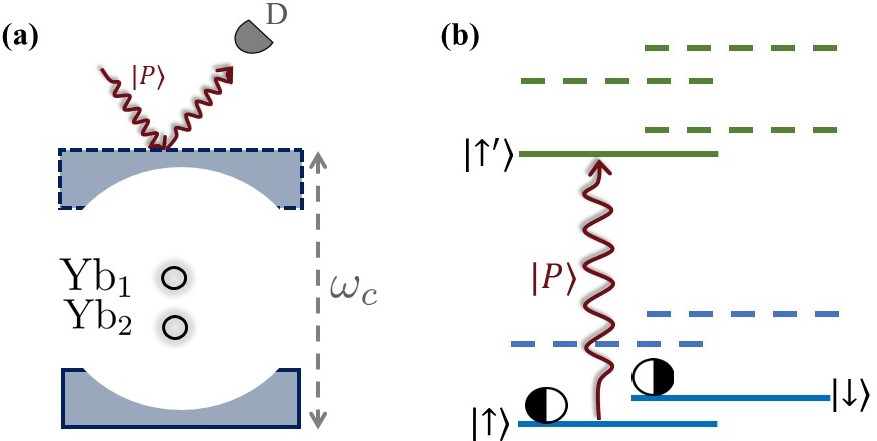}
\caption{ (a) Schematic of the photon interference-based gate scheme setup: Each Yb ion is positioned within a cavity. The spontaneously emitted photons are routed to a 50:50 beam splitter for detection. (b) The hyperfine energy level structures of Yb:YVO taken from \cref{Yb structure}. To execute the gate, a pulse $P_1$,  tuned to resonance with the $\ket{\uparrow} \mapsto \ket{\uparrow'}$ transition, drives the Yb ions into their excited state. Next, a microwave $\pi$ pulse $P_2$, is applied to flip the qubit states (details are discussed in the main text). Subsequently, a second optical excitation is applied using the $P_1$ pulse.} 
\label{barret-kok scheme}
\end{figure}

For this scheme, a three-level system is required, consisting of two qubit levels in the ground state hyperfine manifold, denoted by $\ket{\uparrow}$ and $\ket{\downarrow}$, and an ancillary excited state level, denoted by $\ket{\uparrow'}$, as illustrated in \cref{barret-kok scheme}.  Both ions are pumped to $\ket{\uparrow}$, then a microwave $\pi/2$ pulse creates the superposition $\left(\ket{\uparrow}+\ket{\downarrow}\right)/\sqrt{2}$. The two-qubit operation starts by applying optical $\pi$ pulses in resonance with the $\ket{\downarrow} \mapsto \ket{\uparrow'}$ transition for each ion. The excited states will eventually decay (with potential enhancement from Purcell effects when ions are placed in cavities), creating entanglement between the qubit state and the photon number. This results in the state $\left(\ket{\uparrow0}+\ket{\downarrow1}\right)/\sqrt{2}$, where $1(0)$ indicates the presence (absence) of a photon.
The emitted photon(s) are then sent to a beam splitter (BS) positioned midway between the ions, where the which-path information is erased before the photon detection event occurs. Detection of a single photon will project the joint state of the two ions into a maximally entangled state. However, there is a possibility that both ions were excited and emitted photons, but one of the photons was lost during transmission. To rule out this scenario, which would result in a product state rather than a Bell state, a $\pi$ microwave pulse (pulse $P_2$ in \cref{barret-kok scheme}) is applied to the ground state levels right after the first excitation-emission step, causing a spin flip. This is followed by a second round of optical excitation. Detecting two sequential single photons ensures that the remote Yb ions are in an entangled Bell state
\begin{equation}
 \left| \psi_{\pm}  \right\rangle  = \frac1{\sqrt2} (\left|  \uparrow_{\mathrm{Yb}_1} \downarrow_{\mathrm{Yb}_2} \right\rangle  \pm \left| \downarrow_{\mathrm{Yb}_1} \uparrow_{\mathrm{Yb}_2} \right\rangle ),
\end{equation}
where $+ (-)$ sign indicates the case such that the same (different) detector(s) registered a photon.
It can be shown that by altering the measurement basis, this scheme can be transformed into a CZ gate (see \cref{barret-kok appendix} for the derivation).

\section{Fidelity Computation}\label{sec:state-fid}

In this section, we introduce a perturbative approach to compute the state fidelities' deviation from $1$, serving as a measure of the distance between an output state produced by the imperfect gate to an output state produced by the ideal implementation. In doing so, we fix an input state to the gate. More concretely, letting $\Phi$ denote the noisy implementation of a unitary $U$, we calculate
\begin{align*}\label{eq:state-fidelity}
F(U \ket{\psi_{\mathrm{in}}} \bra{\psi_{\mathrm{in}}} U^\dagger, \Phi(\ket{\psi_{\mathrm{in}}} \bra{\psi_{\mathrm{in}}}).
\end{align*}
We note that in quantum communication applications, we are most interested in generating Bell pairs (or states that are locally equivalent to it). Hence, as the gates we consider here prepare a maximally entangled state on $\ket{\psi_{\mathrm{in}}} = \ket{++}$, we calculate the output state fidelities for the input $\ket{++}$. On the other hand, for broader applications, such as fault-tolerant quantum computation, a noisy implementation $U$, should work well on any input state. In such contexts, we are often interested in the calculation of figures of merit that do not depend on the input state, such as the `average gate fidelity.'
In \cref{gate-fids}, we show how one can generalize the tool we introduce here for the calculation of \eqref{eq:state-fidelity} to efficiently calculate the average gate fidelity as well. Furthermore, in \cref{comparison of fids} we compare the average gate fidelity against the state fidelity with $\ket{\psi_{\mathrm{in}}} = \ket{++}$, for one gate scheme, demonstrating that they both follow a similar trend. These results provide a basis for estimating errors in single- and two-qubit operations.

From this point forward, `fidelity' refers to state fidelity unless explicitly stated otherwise to indicate average gate fidelity.

To quantify the state fidelity of a two-qubit gate, we use tools from time-dependent perturbation theory to analytically derive expressions for the lowest-order non-zero error contributions, which for many realistic applications is sufficient to infer the quality of the gate. We model the noisy gate as a Markovian evolution, which allows us to exploit the Gorini, Kossakowski, Lindblad and Sudarshan (GKLS) master equation to describe the gate dynamics. It is worth noting that this approach works for any initial state.
To compare our method with the prior literature, we highlight that the previous works \cite{asadi2020cavity, asadi2020protocols} have focused on solving the master equations to obtain output state fidelities, which can be challenging, particularly as the system size increases. To clarify, denoting the Hilbert space dimension by $d$ (for our purposes $d$ gets as large as $16$), our method only requires matrix multiplications of size $d$, and has a complexity of $O(d^3)$ with system size. In contrast, previous approaches necessitate the exponentiation of an operator of size $d^2$, representing the noisy Lindbladian superoperator, which translates to $O(d^6)$ computational complexity.

In deriving expressions for the dominant error terms, we consider that imperfections may arise in three different ways: (1) Hermitian perturbations to the gate Hamiltonian, (2) weak Markovian decoherence due to irreversible processes captured by standard Lindblad collapse operators, and (3) non-Hermitian perturbations to the ideal gate Hamiltonian. The first error process gives rise to errors that maintain state coherence and could be eliminated through improved quantum characterization and control. The second error process captures the competition between the desired quantum evolution of the system and its direct interaction with the environment. The third error process can arise due to a combination of the previous two in an analytically simplified model. This latter process is useful for capturing dispersive interactions in the presence of environmental noise, such as in the virtual photon exchange gate \cite{asadi2020cavity} (we refer the reader to \cref{fid:virtual} for the details).

Consider that the desired gate is implemented by a gate Hamiltonian denoted by $H_g(t)$. We assume that this Hamiltonian produces a unitary propagator $U_g(T_g, 0) = \mathscr T\operatorname{exp}(-\frac{i}{\hbar}\int_{0}^{T_g}H_g(t)dt)$ that evolves the quantum system state $\ket{\psi(t)}$ from time $t=0$ to some finite gate time $T_g$, at which time the state $\ket{\psi(T_g)}$ is the ideal result of the gate i.e., the solution to the Schr\"odinger equation
\begin{align}
\frac{\mathrm d}{\mathrm dt} \ket{\psi(t)} = -\frac{i}{\hbar}
H_g(t) \ket{\psi(t)}.
\end{align}

Due to the presence of imperfections, we consider that the actual evolution of the quantum state is governed by a master equation of the form $\frac{\mathrm d}{\mathrm dt} \rho = \mathcal{L}\rho$ for a Lindbladian superoperator $\mathcal{L}$ that is defined by
\begin{equation}\label{eq:lindbladian}
\mathcal{L} = -\frac{i}{\hbar}\mathcal{H} + \sum_k \gamma_k \mathcal{D}(L_k).
\end{equation}
where $\mathcal{H}\rho = H(t)\rho - \rho H^\dagger(t)$ is the Hamiltonian superoperator corresponding to Hamiltonian $H(t) = H_g(t)+\delta \tilde H_e(t)$ for a possibly non-Hermitian perturbation $\tilde{H}_e(t)$ with weight $\delta$. Direct irreversible errors are captured by the dissipative superoperator $\mathcal{D}(L) = L\rho L^\dagger -\frac12\{L^\dagger L, \rho\}$ for all relevant Lindblad collapse operators $L_k$ with associated rate $\gamma_k$.

To simplify the analysis, we make the assumption that the gate time $T_g$ is characterized well and can be implemented with enough precision so that the actual output state is well-approximated by $\rho(T_g)$. That is, we assume errors arising due to other factors will dominate any small differences between $\rho(t)$ and $\rho(T_g)$ when $t\simeq T_g$. In that case, the fidelity $F$ of the actual output state $\rho(T_g)$ with respect to the ideal state $\ket{\psi(T_g)}$ is then given by
\begin{align}\label{eq:fidelity-definition}
F = \bra{\psi(T_g)}\rho(T_g)\ket{\psi(T_g)}
\end{align}
and we define the corresponding error as $\epsilon = 1 - F$. Note that the assumption on $t\simeq T_g$ can be relaxed but the result is a more complicated expression that is less intuitive.

Using time-dependent perturbation theory (see \cref{app:perturbation}), we find that the error can be expanded into
\begin{align}\label{eq:hermitian-infidelity}
\epsilon &= \epsilon^{(1)}_{L} + \epsilon^{(1)}_{H} + \epsilon^{(2)}_{HH} + \epsilon^{(2)}_{LH} + \epsilon^{(2)}_{LL} +\cdots
\end{align}
where the superscript $(n)$ indicates the $n$-th order perturbation to $\epsilon^{(0)}=0$ and the subscripts $H$ and $L$ indicate whether the term arises due to perturbations of $H$ and irreversible processes captured by $L_k$, respectively. The first-order terms are given by
\begin{align}
\epsilon_L^{(1)} &= \sum_{k} \int_{0}^{T_g} \gamma_k \left( \langle L_k^\dagger(t) L_k(t) \rangle - |\langle L_k(t) \rangle|^2 \right) \, \mathrm dt \label{eps-first-order}\\
\epsilon^{(1)}_H &= -\frac{2\delta}{\hbar}\int_{0}^{T_g} \Im\left( \langle \tilde{H}_e(t)\rangle\right) \, \mathrm dt.\label{eq:non-hermit-expression}
\end{align}
Here, we have employed the convention that $\langle\mathcal O(t)\rangle=\bra{\psi(t)}\mathcal O\ket{\psi(t)}$ for an operator $\mathcal O$.

An interesting consequence of the first-order result is that $\epsilon_H^{(1)}$ vanishes if $\tilde{H}_e$ is Hermitian. To obtain the first non-zero term arising from reversible errors (Hermitian perturbations to $H_g$), it is necessary to go to the second-order terms. However, since $\epsilon_L^{(1)}$ is generally non-zero at the first order, it will tend to dominate over $\epsilon_{HL}^{(2)}$ and $\epsilon_{LL}^{(2)}$. The expressions for these two terms are presented in the \cref{app:perturbation} for an interested reader.
Here, out of second order terms, we only focus on the reversible error $\epsilon_{HH}^{(2)}$ due to the Hermitian perturbation $H_e = (\tilde{H}_e + \tilde{H}_e^\dagger)/2$, which can be non-zero even in the absence of irreversible errors
and is given by
\begin{align}\label{eps-second-order}
\begin{split}
\epsilon^{(2)}_{HH} &= \frac{2 \delta^2}{\hbar^2}\int_{0}^{T_g} \!\!\!\int_{0}^{t}\!\operatorname{Re}\bigg(\langle H_e(t) H_e(t')\rangle\\
&\qquad\qquad\qquad\qquad-\langle H_e(t)\rangle\langle H_e(t')\rangle\bigg)\mathrm dt \mathrm dt',
\end{split}
\end{align}
where the two-time correlation is computed by $\langle \mathcal O_1(t) \mathcal O_2(t')\rangle = \bra{\psi(t)} \mathcal O_1 U_g(t,t') \mathcal O_2 \ket{\psi(t')}$ for any pair of operators $\mathcal O_1,\mathcal O_2$.

For the three expressions obtained for
$\epsilon_L^{(1)}$, $\epsilon_H^{(1)}$ and $\epsilon_{HH}^{(2)}$ to be evaluated analytically it is necessary to obtain an analytic solution to the ideal gate evolution. If the integrals can also be computed analytically, we will have a full analytical expression. However, in case the integrals cannot be computed analytically, one can utilize numerical methods to compute the infidelities for a fixed gate time $T_g$ in terms of the imperfection parameters (i.e., $\delta$ and $\gamma_k$). This is generally much easier than solving the full system master equation to obtain the exact fidelity.

In the following, we use this method to compute the state fidelity of the magnetic dipolar gate, which serves as an example of having a Hermitian error Hamiltonian. Additionally, in \cref{fid:virtual}, we re-compute the state fidelity of the virtual photon exchange gate \cite{asadi2020cavity} as an example of a non-Hermitian perturbation Hamiltonian. However, this gate and its fidelity calculation have already been discussed in \cite{asadi2020cavity, asadi2020protocols, wein2021modelling}.

\section{Fidelity estimations and implementations }\label{imp}

In this section, we provide a detailed analysis of the state fidelity expressions for each gate scheme, using both analytical and numerical techniques to evaluate their performance for quantum networks purposes. In our analysis, we include the errors that are considered fundamental to the schemes, and we highlight that technology-related issues, such as collection efficiencies, will ultimately need to be improved.
We also discuss the requirements for the experimental implementation of the parameters involved in the fidelity expressions. Based on these results, we estimate the achievable fidelity of each gate scheme for individual Yb ions at the current state of technology.

\subsection{Magnetic dipolar gate}\label{impl:MD}

As discussed in \cref{sec:magnetic-dipolar}, this scheme involves two sets of qubits (active and passive) defined in the ground and excited states of each ion. Since the two-qubit phase gate is effectively performed in the excited state before the quantum state is transferred back to the ground state, we consider the optical decay rate, optical pure dephasing rate, as well as the ground- and excited-state decay and dephasing rates as the main fundamental limiting factors to be included in the fidelity calculation. In addition, errors arising from the interaction Hamiltonian are also taken into account. In the following, we present the state fidelity of this gate, along with a detailed discussion of several secondary effects that are neglected in the model, as they are either expected to be significantly improved under optimized conditions or negligible compared to the dominant error sources.

Employing the perturbative method, which was described in the previous section, allows us to compute the state fidelity of the magnetic dipole-dipole interaction gate as (see \cref{magnetic} for the derivation)
\begin{align}\label{f_md}
\begin{split}
\mathrm{F_{MD}}=&
1-{T_\mathrm{act}}\bigg(\frac{{7}}{{8}}({\gamma _{1, \Uparrow '}}  + {\gamma _{1, \Downarrow '}}) +\frac{1}{2}(\gamma_{2,\Uparrow'}+\gamma_{2,\Downarrow'})\\
&\qquad\quad +\frac{13}{16}({\gamma _3} + {\gamma _4}) + \frac{1}{2}({\gamma _5} + {\gamma _6})\bigg)  \\ 
 &-{T_{{\mathop{\rm int}} }}\left({\gamma _{1, \Uparrow '}} + {\gamma _{1, \Downarrow '}} + \frac{{3}}{{4}}{\gamma _4} + \frac{{1}}{{2}}{\gamma _6}\right) \\
 &-\frac{a\left(J_x + J_y\right)^2}{{J_{z}^2}},
 \end{split}
\end{align}
where $T_{act} = \frac{\pi}{\Omega}$ represents the activation time, the duration needed to transfer the population from passive to active qubits, with $\Omega$ being the Rabi frequency.
Here the dissipative parameters affecting the system are considered as optical decay rate $\gamma_{1,\Uparrow'}(\gamma_{1,\Downarrow'})$, optical pure dephasing rate in the bulk $\gamma_{2,\Uparrow'}(\gamma_{2,\Downarrow'})$  corresponding to the transitions $\Uparrow':\ket{\uparrow'}\mapsto\ket{\uparrow}$ ($\Downarrow':\ket{\downarrow'}\mapsto\ket{\downarrow}$), ground (excited) state spin decay rate $\gamma_{3(4)}$, and ground (excited) state spin dephase rate $\gamma_{5(6)}$.
The interaction time between two active qubits in the excited states is denoted by $T_{int}=\frac{\hbar\pi}{4J_{z}}$ where $J_{z}=\frac{\mu _0(\mu _Bg_{z})^2}{8\pi r^3}$ is the coupling strength between the qubits, with $\mu_0$ being the vacuum permeability, $\mu_B$ the Bohr magneton, $g_z$ as the principal value of the g-tensor, and $r$ the spatial distance between two Yb ions.
In this analysis
$J_i=\frac{\mu_0(\mu_Bg_i)^2}{16\pi r^3}$ (for $i=x,y$) define the transverse
components of the dipolar interaction, where we consider the point symmetry with $g_x=g_y \equiv g_{\perp}$ and $g_z\equiv g_{\|}$. Here, $a$ is a coefficient defined by $a = (32(2-\sqrt{2}) - {\pi ^2})/64$.
For $J_x=J_y \equiv J_\perp$ the second order error (denoted by coefficient $a$) is bounded above by $\sim J_\perp^2/J_{\|}^2$ which improves the estimation obtained in Ref. \cite{grimm2021universal}. 
In ${}^{171}$Yb$^{3+}$:YVO, the excited state g-tensor components parallel and perpendicular to the crystal symmetry axis (the \(c\) axis) are \(g_{\|} = 2.51\) and \(g_{\perp} = 1.7\) \cite{kindem2018characterization}. This results in a second-order error contribution to the fidelity of approximately \(10^{-2}\), setting a lower bound for the fidelity. This lower bound is relevant in cases where transverse interactions mediate the gate process.

The optical decay rate for an ensemble of ${}^{171}$Yb$^{3+}$: YVO is $\gamma_{1,\Downarrow'}\simeq\gamma_{1,\Uparrow'}\equiv\gamma_1=2\pi\times596$ Hz \cite{kindem2018characterization}.
The optical pure dephasing rate in the bulk is estimated with relation $\gamma_{2,\Uparrow'}\simeq\gamma_{2,\Downarrow'}\equiv\gamma_2=1/T_{2o}-1/2T_{1o}\approx9.12$kHz, where $T_{2o}=91\mu$s is the optical coherence time of an ensemble of Yb ions in YVO bulk at B$\approx500$ mT, and $T_{1o}=267\mu$s in the optical life time of Yb ions in YVO bulk \cite{kindem2018characterization}. We estimate the ground state spin decay rate $\gamma_3=1/T_{1s,g}$, where $T_{1s,g}$ is the ground state spin relaxation time and was measured to be longer than $200$ ms \cite{macfarlane1987coherent, kindem2018characterization}. We assume that the excited state spin decay rate ($\gamma_4$) is the same as the ground state ($\gamma_3$), as there is no available information on this to the best of our knowledge.
We also estimate the ground state spin dephasing rate with relation $\gamma_5=1/{T_{2s,g}}-\gamma_3/2\approx149$ Hz, where $T_{2s,g}=6.6$ ms (at B=440 mT) is the ground state spin coherence time for an ensemble of Yb ions in YVO bulk\cite{kindem2018characterization}.
Similarly, we estimate the excited state spin dephasing rate with relation $\gamma_6=1/T_{2s,e}-\gamma_4/2\approx28.57$ kHz, where $T_{2s,e}$ is the excited state spin coherence time,  measured to be $35\mu$s at low magnetic fields \cite{bartholomew2020chip}. However, to the best of our knowledge, this parameter has not been measured for single Yb:YVO ions at higher magnetic fields.
The overall gate time is defined to be $T_{g,MD}=2T_{act}+T_{int}$, where the coefficient $2$ is considered for activation and deactivation pulses. 

\begin{figure}
\centering
\includegraphics[width=0.5\textwidth]{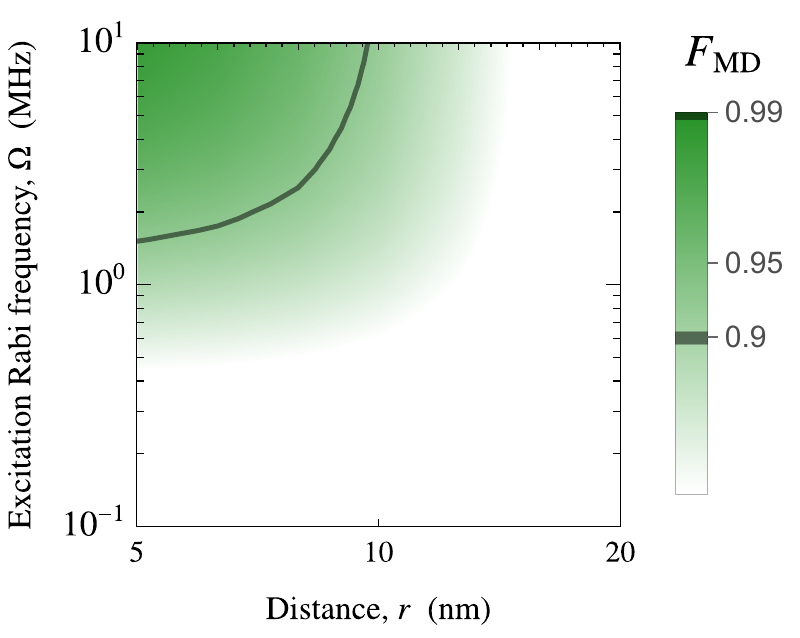}
\caption{Fidelity of the magnetic dipolar gate as a function of Rabi frequency (activation pulse) and the distance between two Yb ions doped into YVO crystal. If two ions are spatially far away more than $10$ nm, fast excitations may still achieve a fidelity of above $90\%$ for the magnetic dipolar gate. Here, we start the plot at 5 nm, where modelling the system through the perturbation method (i.e., \cref{f_md}) matches the exact simulation of the system (see \cref{MD-state-fids} for more details). The maximum fidelity achieved in this plot is $\approx 0.95$. The parameters used in the simulation for this figure are provided in this subsection. They are also summarized in \cref{tab:YbYVO_parameters}.}
\label{imp:MD}
\end{figure}

Given that the atomic properties of the system are fixed, as stated in \cref{f_md}, the only experimental parameters we can adjust to achieve high fidelity are the ion separation and the laser power. Therefore, if one of these parameters is constrained, it may be practical to vary the other to explore conditions that yield good performance. For example, if the splitting is small, achieving fast excitations can be challenging.
In that case, one can examine how close the ions have to be to obtain decent fidelities. Therefore, there is a trade-off between splitting and separation of ions. In \cref{imp:MD}, we demonstrate fidelity as a function of Rabi frequency and ions' separation, illustrating the variation in laser power corresponding to the changing distance between two ions.

We note that among the schemes discussed in this paper, the magnetic dipolar interaction–based approach can be implemented without a cavity but still requires an external magnetic field to split the hyperfine levels. Accordingly, we primarily relied on reported coherence times in YVO bulk under nonzero magnetic fields. These data are listed in \cref{tab:YbYVO_parameters}. However, most of these data correspond to ensembles of Yb ions rather than single ions. Since single-ion systems typically exhibit longer coherence times due to the absence of the ensemble-induced flip-flop mechanism, our fidelity estimates likely represent a lower bound on the achievable fidelity once single-ion data become available.

In modelling the system, we neglect certain mechanisms. Below, we describe how these effects scale relative to the two-qubit process.
We note that \cref{f_md}, and hence the $1-O(r^3)$ fidelity scaling, is obtained under the assumption that direct interaction is negligible during excitations. More formally, we assume $\norm{H_{\mathrm{int}} T_{\mathrm{act}}} \ll 1$ with $H_{\mathrm{int}}$ being the dipolar interaction Hamiltonian. However, when considering $\Omega=10$MHz, the numerical simulation of the Yb system shows that this assumption breaks down at small separations with a turning point around $r\sim 5$ nm. Therefore, for Yb ions, we anticipate the highest fidelity around this separation. See \cref{magnetic} and \cref{MD-state-fids} for more details. For $r>5$ nm, as \cref{imp:MD} confirms, the distance cannot be arbitrarily large, as the fidelity of the two-qubit process decreases with increased separation.
Two nearby ions can be found by implanting each ion at precise locations within the crystal, achieving a separation of just a few nanometers between them \cite{kornher2016production}.

Since the magnetic dipolar gate-based scheme relies on direct qubit–qubit interactions, its performance is constrained by the spatial separation between the ions and requires them to be in close proximity. However, if the ions are too close, applying a single-qubit gate to an individual ion can inadvertently excite nearby ions—particularly in the microwave domain. Therefore, for this scheme to be functional, the ions must be spatially resolved so that a single ion can be addressed without influencing its neighbours. Moreover,  if one uses microwave pulses for the single-qubit gate operation, spectral addressing is required as well.
For example, a field gradient can be applied to detune the target ion from nearby ions, allowing it to be spectrally addressed at a specific time.
As shown in \cite{grimm2021universal}, instead of microwave pulses, one can take advantage of fast optical pulses to implement the single-qubit gates. In the optical domain, closely spaced ions can have distinct transition frequencies due to inhomogeneous broadening.
In this case, the single-qubit gate time can be estimated as  $T_{\rm single} = \frac{\pi}{\Omega},$
where $\Omega$ is the Rabi frequency of the transition.   
For example, for a Rabi frequency of $\Omega=10\,\text{MHz}$, the activation time is about $T_{\rm act}=\pi/\Omega=0.31\,\mu\text{s}$ and the single-qubit Hadamard gate (implemented through two consecutive $\pi$ pulses) takes $T_{\rm single}=2\pi/\Omega=0.62\,\mu\text{s}$.

Following Ref. \cite{grimm2021universal}, we disregard the nuclear magnetic dipolar interaction, as its strength is $\sim 10^7$ times weaker than the electronic magnetic dipolar interaction. This difference arises mostly from the small nuclear magnetic moment ($\mu_N$), and the nuclear g factor ($g_n=0.987$).

When performing the CZ gate, some infidelity inevitably arises from the single-qubit rotations.
Ref. \cite{grimm2021universal} discusses that the intrinsic optical decoherence rate leads to an infidelity of $\varepsilon_{\rm single} \sim \frac{T_{\rm single}}{T_{2o}}$ where $T_{2o}$ is the optical coherence time. Using $T_{2o}\approx91\mu\mathrm{s}$ (at B = 500 mT) \cite{kindem2018characterization}$,$ and the aforementioned practical single-qubit gate time, the resulting error is $\varepsilon_{\text{single}} \sim 10^{-3}$, which is negligible.
However, reducing the Rabi frequency in \cref{imp:MD} increases the gate time and, in turn, the single-qubit gate error. To address the single-qubit gate error scaling, we note that single-qubit gate errors were not included in our analysis, even for small Rabi frequencies.   
The main reason is that for the regimes in \cref{imp:MD}, i.e., $\Omega=10,\,1,\,0.1\,\text{MHz}$, the single-qubit gate errors are $10^{-3},\,10^{-2},\,10^{-1}$, respectively.
This indicates that increasing the Rabi frequencies will reduce the corresponding error.
These errors remain at least one order of magnitude smaller than the two-qubit gate errors for the regimes where our fidelity equation (obtained through the perturbative approach) is valid (green area in \cref{imp:MD}). 
Moreover, the excitation pulses are considered to be perfect.
Therefore, in these analyses, we assume that the fidelity of the CZ gate is primarily determined by the two-qubit gate processes. However, we include these fast excitations in our analysis to capture their effect on the two-qubit activation and diactivation process (see the second term in \cref{f_md}).\\

While exciting the Yb ions, there are two kinds of off-resonant excitations. The first is off-resonant excitation due to optical $\pi$ pulses (involved in both single-qubit and activation operations) from other energy levels of the target Yb ion (qubit), which causes infidelity of  $
\sim \exp\!\left(-\frac{\pi}{2}\left(\frac{\Delta_s}{\Omega}\right)^2\right)$,
with $\Delta_s$ being the hyperfine splitting~\cite{grimm2021universal}. To avoid any off-resonant excitation, the Rabi frequency of the activation pulses should be much smaller than the smallest hyperfine splitting. Choosing $\Delta_s \geq 2.5 \Omega$ yields an infidelity of $\sim 10^{-4}$.  Thus increasing the $\Delta_s/\Omega$ ratio (i.e., using small Rabi frequencies) decreases errors from off-resonant excitation. One can easily see that a trade-off exists between errors from single-qubit gates and those from off-resonant excitations of other Yb qubit energy levels: decreasing the Rabi frequency reduces off-resonant errors but increases single-qubit gate errors.
The second type is off-resonant excitation of non-target Yb ions. To suppress such excitations, as discussed in~\cite{grimm2021universal}, electrical gating can be applied to locally shift the energy levels of the neighboring Yb ions.

In this gate scheme, the activation process is separate from the two-qubit interaction that performs the gate. Consequently, no AC Stark shift occurs during the interaction itself, as the laser is turned off at that stage.
The AC Stark shifts arise only during the qubit activation process.  
More specifically, when applying the $\pi$-pulses to activate the qubits, the relative AC Stark shift of the two transitions may induce a relative phase on the qubit. This effect can be regarded as a local single-qubit gate (i.e., a coherent phase error). In other words, the spin levels in the excited state acquire a relative phase determined by the power used for the two $\pi$-pulses on each transition, as each transition may experience a slightly different AC Stark shift.  
We note that coherent errors are not included in our model, and the fidelity expression accounts only for dissipative errors. However, we recognize that when implementing this gate, one must characterize and track the actual phase accumulated during excitation. Furthermore, upon inverting the system, it is necessary to identify the appropriate power for the $\pi$-pulses that return the population to the ground state (di-excitation). Even under this condition, additional relative phases between the spin states could be accumulated. These should be characterized and compensated with local corrections when necessary. 
Nonetheless, we note that this effect could manifest as a stochastic error if the laser is not sufficiently stable, as power fluctuations would cause the relative phase to vary over time. However, this lies beyond our stable-laser assumptions used in our work.

For the schemes studied in this work, spectral diffusion is not included in the simulations. 
The main reason is that we assume that the gate operations are sufficiently fast for the ions to effectively avoid experiencing spectral diffusion during this period.

As discussed earlier in the paper, the potential advantages of the magnetic dipole gate include the fact that it is a deterministic scheme, as well as the fact that it does not directly require a cavity for its operation. However, for other parts of the network—such as initialization and qubit readout—a cavity may still be needed. 
One possible solution is to use the transitions that are off-resonance with the cavity as qubit states during gate operation, while bringing other transitions into resonance with the cavity for qubit readout. Alternatively, an electric field gradient could be applied to tune the relevant transitions into resonance with the cavity for initialization, and then removed for gate operation.

To limit the decay process, we choose two of the lowest ground state hyperfine levels with long enough coherence time as the passive qubit levels.
As discussed in \cref{sec:magnetic-dipolar}, we employ optical pulses to define the active qubits in the excited states.
Furthermore, in ${}^{171}$Yb$^{3+}$:YVO system with a nonzero perpendicular component of the g-tensor $(g_{\perp}\ne0)$, choosing opposite nuclear levels within each electronic doublet reduces spin flip-flop errors during the two-qubit gate process \cite{grimm2021universal}. This approach offers two options for selecting active qubits within both different electron spin and nuclear spins (e.g., both solid or dashed green lines in \cref{yb-qubits}).
An alternative approach is to define both passive and active qubits in the ground state. In this configuration, passive qubits occupy the $m_s = -1/2$ manifold (with qubit states $m_I = \pm 1/2$), while active qubits reside in the $m_s = +1/2$ manifold (also with $m_I=\pm 1/2$). However, implementing active qubits in the excited state via optical transitions offers clear advantages for realizing this gate in the Yb:YVO system and, more generally, in other rare-earth platforms with long optical lifetimes~\cite{grimm2021universal}. Optical excitation enables a faster activation process than ground-state microwave driving and simultaneously suppresses unwanted microwave coupling to nearby (non-qubit) ions. 
Moreover, when qubit states belong to different electron spin manifolds, the dipolar interaction strength is mediated by the Bohr magneton $\mu_B$, whereas if both qubits lie within the same $m_s$ manifold, the coupling is much weaker, being governed by the nuclear magneton $\mu_N = \mu_B / 1836$.
\\

\subsection{Photon scattering gate}\label{ps implemnt}

In this scheme, photon loss due to spontaneous emission, scattering, detector efficiency and other deficiencies make the gate operation probabilistic. However, they do not cause errors as long as the photon count is properly recorded \cite{duan2005robust}. Therefore, we assume that the photon is successfully detected and focus on the remaining sources of error that impact the fidelity. The fidelity of the photon-scattering gate scheme in high cooperativity regime $C\gg1$ is computed in \cite{asadi2020cavity, c8yf-kv59} to be

\begin{equation}\label{fid:PS}
\begin{aligned}
\mathrm{F_{PS}} =& 1 - 
\frac{11}{16C^2} - \frac{(\delta_p^2+\sigma_p^2)}{4\gamma_1^2 C^2}\left( 11 - 80\alpha^2 + 192 \alpha^4 \right)\\&
-\frac{41\delta_{Yb1}^2-38\delta_{Yb1}\delta_{Yb2}+41\delta_{Yb2}^2}{16\gamma_1^2C^2}\\&
-\frac{(-11+40\alpha^2)(\delta_{Yb_1}+\delta_{Yb_2})\delta_p}{4\gamma_1^2C^2}
-\Gamma T_{g,PS}, 
\end{aligned}
\end{equation}
where $\delta_p$ is the cavity-photon detuning, $\delta_{Yb_n}$ denotes the detuning of the cavity frequency from the optical transition of the 
$n^{\text{th}}$ system, $\sigma_p$ is the spectral standard deviation of the incident photon that is modelled assuming a Gaussian wave packet for the photon, $\Gamma$ represents the effective decoherence rate, and $T_{g,PS}$ is the gate time. The parameter $\alpha = g/\kappa$ defines the ratio between the cavity-ion coupling rate $g$, and the cavity decay rate $\kappa$ and
the cavity cooperativity is $C=4g^2/\kappa\gamma_1$.

As discussed in \cref{Duan-Kimble}, in this scheme the qubits are defined in the ground-state levels of each ion, while a single auxiliary excited state is addressed to mediate the gate interaction. Consequently, the main fundamental limiting factors are expected to be the optical decay rate, optical pure dephasing and spin decoherence of the Yb$^{3+}$ ions within the cavity. In addition, the incoming photon and cavity parameters—such as the cooperativity, the ratio of cavity coupling to cavity bandwidth, and the photon bandwidth—play crucial roles and must be included in the fidelity analysis. In the following, we present our approach for incorporating these parameters into the gate simulation. We also discuss how certain other parameters introduced in \cref{fid:PS} can be neglected to simplify the simulations.

While the analytical expression for gate fidelity has been previously derived, it does not account for optical pure dephasing and spin decoherence effects. To address these limitations, we approach the problem using a fully numerical method based on input-output theory (see \cref{app: photon scattering}). However, in our numerical simulation, we consider the photon to have a Lorentzian spectrum rather than the Gaussian spectrum used in the analytic expression since otherwise a large time-dependent master equation must be solved that substantially increases simulation time and restricts the parameter range that we can explore. Thus, although the qualitative behaviour and scaling properties are expected to be the same as in \cref{fid:PS}, quantitative comparisons require caution.

\begin{figure}[!t]
\centering
\includegraphics[width=0.5\textwidth]{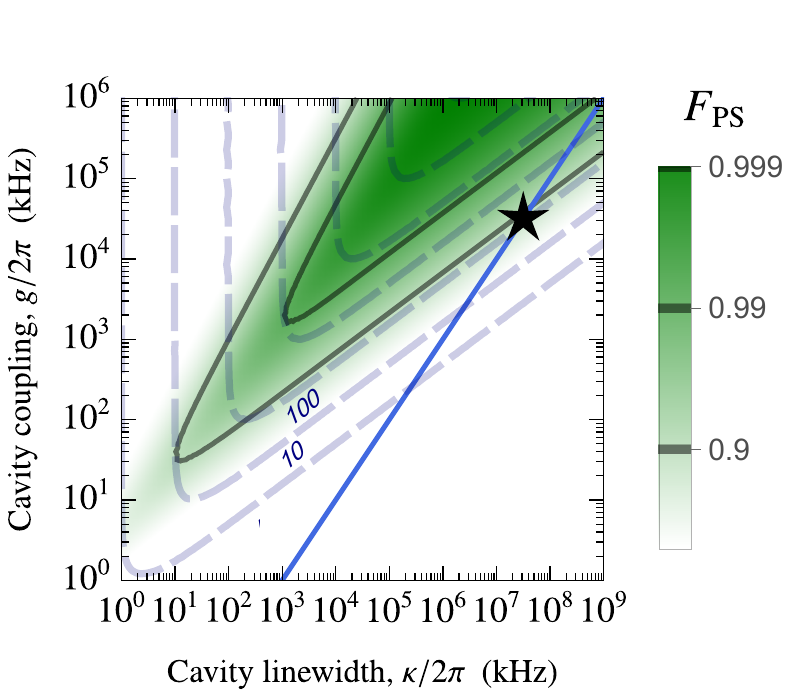}
\caption{Fidelity of the photon scattering gate as a function of cavity parameters. 
The dashed lines represent the necessary maximum Purcell enhancement (when on resonance) for each regime, increasing exponentially by powers of ten ($10,10^2,10^3$, and so on).
Here we assume $\delta_p=\delta_{Yb_1}\!-\delta_{Yb_2}=0$. The cavity coupling and decay rates set to $g = 2\pi \times 23$ MHz, and $\kappa = 2\pi \times 30.7$ GHz \cite{kindem2020control}. Since the presence of a cavity can negatively impact the optical coherence, we use a slightly reduced coherence time of $T_{2o}\approx39 \mu$s compared to the bulk value of $T_{2o}\approx91\mu$s (at B$\approx500$mT) \cite{kindem2018characterization} prior to accounting for Purcell enhancement, which we estimated from Ref. \cite{kindem2020control}. The parameters used for this figure are described later in this subsection and are summarized in \cref{tab:YbYVO_parameters}. 
We also note that dark counts are neglected in our gate modeling, assuming that moderate detection efficiencies can be achieved (see \cref{implementation:BK scheme} for further discussion).
The star mark indicates the current achievable fidelity of this scheme for Yb:YVO to be 0.8. The realistic parameter regimes achievable by a physical device are expected to center around the star mark (reflecting current implementations), whereas reaching the strong coupling regime remains highly challenging.
The blue line indicates the current ratio $\alpha=g/\kappa$ and is studied separately in \cref{comparision3}. } 
\label{imp:PS}
\end{figure}

In \cref{imp:PS}, we display our fidelity estimations as a function of cavity parameters ($g,\kappa$) using the numerical simulation where we consider a photon with a Lorentzian-shaped spectrum. 
We also use this numerical simulation to validate properties of the previous analytical approximation derived in \cref{fid:PS}.
Our numerical simulation reveals that we can capture the effect of optical pure dephasing by subtracting $\approx(11\gamma^{\ast})/(8\gamma_1C)$. 
Additionally, we find that 
$\Gamma \propto 1/T_{2s,g,\mathrm{cav}}$, where $T_{2s,g,\mathrm{cav}}$ represents the ground state spin coherence time in cavity, offers a reliable estimate of the error caused by spin decoherence. The exact proportionality coefficient, however, depends on the shape of the photon.
When using the same definition of gate time as in \cite{asadi2020cavity} along with $\sigma_p=1/T_{1p}$ where $T_{1p}$ is the scattered photon lifetime, we find $\Gamma \simeq 1/T_{2s,g,\mathrm{cav}}$ quantitatively agrees with the numerical simulation.

The numerical simulation also leads us to identify a previously-overlooked alteration of the lower-bound error scaling when spin decoherence is non-negligible. 
Assuming $\delta_p=0$ and the ions are in perfect resonance with each other and with the cavity, from \cref{fid:PS}, we find that in the absence of spin dephasing ($\Gamma=0$) and optical pure dephasing ($\gamma^{\ast}=0$), it is always possible to reach the minimum error of $1 -\mathrm{F_{PS,max}}\propto 1/C^2$
by scattering a very narrow photon ($\sigma_p/\gamma \ll 1$). However, the gate time is proportional to the time it takes to scatter the photon ($T_{g,PS}\propto 1/\sigma_p$). Thus, as $C$ is increased to improve fidelity, it is necessary to increase $T_{g,PS}$ and so if $\Gamma \ne 0$ this causes the last term in \cref{fid:PS} to increase. This imposes a trade-off between having an error from the spectral broadening (i.e., $\sigma_p^2$ term) or from the spin dephasing (i.e., $\Gamma T_{g,PS}$ term). In Ref. \cite{asadi2020cavity}, this trade-off was explored but assumed to always be of a second-order correction to the seemingly dominant $1/C$ term when the gate time was optimized. This is, in fact, not the case.

For a given cavity cooperativity and spin dephasing rate $\Gamma$, there is an optimal bandwidth of the photon that minimizes the error. This optimal bandwidth can be computed by setting the third and the last terms in \cref{fid:PS} to be equal. Solving for the bandwidth implies that it must scale as $\sigma_p\propto C^{2/3}$ and hence $\Gamma T_{g,PS}\propto 1/C^{2/3}$. Thus, if either $\Gamma$ or $C$ is large enough, this scheme will have an error scaling of $1/C^{2/3}$ rather than $1/C$, which is significantly worse for large $C$. In our simulations, we numerically optimize the spectral standard deviation ($\sigma_p$) to achieve maximum fidelity.

In addition to setting the photon bandwidth, it is necessary to define a photon truncation time that defines the total gate time (denoted by $T_{g,PS}$). If one stops measuring the photon too soon, there will be low efficiency. On the other hand, if one waits too long, spin decoherence will be exacerbated. For the value of $T_{2s,g,\mathrm{cav}}$ expected, we select this additional truncation time to be proportional to the time-scale of the scattered photon and such that it has a minimal impact on the fidelity for the range of cooperativity explored, leading to $T_{g,PS}=7T_{1p}$ which implies only 0.1\% loss.

Photonic cavities have been fabricated in YVO crystals using ion beam milling, achieving a cavity-Yb ion coupling rate of $g = 2\pi \times 23$ MHz and a cavity decay rate of $\kappa=2\pi \times 30.7$ GHz \cite{kindem2020control}.  
For single Yb ions coupled to this cavity, a reduced optical lifetime of $T_{1o,\mathrm{cav}}=2.27\mu$s has been measured \cite{kindem2020control}.
These are the parameters used in our simulations.
However, the fabrication process poses a significant risk of crystal damage, making it a challenging method. An alternative approach involves designing a hybrid photonic crystal cavity system, where the cavity is fabricated separately and later transferred onto the crystal.
For instance, a Gallium Arsenide (GaAs) photonic crystal cavity has been developed, demonstrating a reduced optical lifetime of $4.2 \mu$s corresponding to the Purcell factor of 179  for strongly coupled single Yb ions coupled to hybrid GaAs-YVO system \cite{wu2023near}.

The zero-first-order Zeeman (ZEFOZ) transitions, which exhibit first-order insensitivity to magnetic field fluctuations, can significantly extend spin coherence. 
Interestingly, Yb:YVO exhibits a ZEFOZ transition (also called clock transition) at B=0, which makes this regime particularly favourable. Therefore, for photon scattering and photon interference-based schemes, for which the performance does not depend on an external magnetic field, we focus on B$=0$. 
For single $^{171}\text{Yb}^{3+}$ ions coupled to a YVO photonic crystal cavity, the ground state spin coherence time is measured to $T_{2s,g,\mathrm{cav}}=31$ ms using Carr-Purcell Meiboom-Gill (CPMG) decoupling sequences and the ground state spin lifetime is measured to be $T_{1s,g,\mathrm{cav}}=54$ ms \cite{kindem2020control}.

To estimate the optical pure dephasing rate in the cavity ($\gamma^\ast$), for photon scattering and photo interference-based schemes,
we use the reduced optical lifetime and optical coherence time measured in the cavity. Therefore, we utilize the relation $\gamma^{\ast} = 1/T_{2o,\mathrm{cav}} - 1/(2T_{1o,\mathrm{cav}})$ to account for the effect of optical pure dephasing arising from charge fluctuations. 
Using $T_{1o,\mathrm{cav}} = 2.27 \, \mu\mathrm{s}$ and $T_{2o,\mathrm{cav}} = 4.1 \, \mu\mathrm{s}$ from \cite{kindem2020control}, 
we obtain an optical pure dephasing rate of $\gamma^{\ast} = 23.64 \, \mathrm{kHz}$. The optical coherence time of a single ion in a cavity is shorter than in bulk. 
This reduction arises from both the shortened optical lifetime in the cavity due to Purcell enhancement and additional dephasing mechanisms in the cavity, 
such as charge fluctuations and fabrication-induced disorder. We also distinguish between the optical coherence time and the measured Ramsey coherence time of $T_{2}^{\ast} = 320 \, \mathrm{ns}$ \cite{kindem2020control} as the latter includes contributions from very slow noise leading to spectral diffusion, 
which we neglect in all gate schemes.

Similar to the magnetic dipolar–based scheme, we assume the gate is fast enough for spectral diffusion to be negligible in the simulation. 
Here, we focus on the upper bound fidelity
 imposed by the measured optical coherence time value only. Although the
 spectral diffusion can be overcome with additional filtering or
 time binning, if unmitigated, this effect can be incorporated
 into the fidelity calculation. Assuming that spectral diffusion
 primarily determines the detuning between the ions’ optical
 transitions during the gate time, one can approximately
 account for this by including the time-dependent spectral
 diffusion function in the fidelity equations via the $\delta_{Yb_1}\!+\delta_{Yb_2}$ term.

In our simulations, we assume that the excitation pulses are perfect, allowing us to restrict our analysis to the interaction process. In this scheme, similar to the previous approach, we employ two hyperfine levels in the ground state as qubit levels. However, unlike the previous scheme, only a single hyperfine level in the excited state is required to serve as an auxiliary state.
After qubit initialization,
the transitions $\ket{\uparrow} \mapsto \ket{\uparrow'}$ of each ion should be brought into resonance with one another and with the cavity mode. 
A magnetic field gradient can tune the optical transitions of the ions into resonance with each other.
On the other hand, to bring the cavity-ion system into resonance, the piezoelectric effect can be employed to adjust the cavity frequency relative to the transition frequency of the ions within it \cite{goswami2018theory}.
When bringing the ions into resonance with each other, it should be noted that the laser field can couple both hyperfine qubit levels to the excited state. To avoid this this undesired coupling, the transition $\ket{\downarrow} \mapsto \ket{\uparrow'}$ should be detuned far from the cavity resonance. However, the currently available cavity designed for this system has a linewidth of $\kappa = 2\pi \times 30.7$ GHz \cite{kindem2020control}, which is relatively wide and effectively resonant with all hyperfine transitions within the $^2F_{5/2}(0) \leftrightarrow {}^2F_{7/2}(0)$ manifold. 
Nevertheless, some transitions are forbidden, while some are suppressed by polarization selection rules, and some are far off-resonant for the narrow-bandwidth incoming photon, whose bandwidth $\sigma_p$ is on the order of kHz to MHz.
Furthermore, introducing a Purcell factor can enhance the cyclicity of the desired transition \cite{kindem2020control}. Together, these elements make the implementation of this scheme possible.
\\

\subsection{Photon interference-based gate}\label{implementation:BK scheme}

In this scheme, the fidelity is primarily restricted by the optical quality of the source. Specifically, when transferring to an MUB basis, we assume all single-qubit gates and the beam splitter to be ideal. 
Excitation pulses are also considered to be perfect.
Therefore, we consider the fidelity of the two-qubit gate to be limited by the two-qubit operation. Since this scheme operates faster than the photon-scattering approach, the primary fundamental limiting factors are the optical decay and pure dephasing rates. In the following, we discuss why other parameters can be neglected.\\

This scheme requires the detection of two consecutive single photons. 
Assuming the detection time ($T_d$) is longer than the optical lifetime ($T_d> T_{1o}$) the fidelity is given by \cite{wein2020analyzing, asadi2020protocols}%
\begin{equation}\label{f_bk}
\mathrm{F_{PI}} = \frac{1}{2}\left(1 + \frac{\gamma_1 ^2}{(\gamma_1+2\gamma_2) ^2+\Delta_\omega^2}\right),
\end{equation}
where $\gamma_1$ represents the optical decay rate from the auxiliary excited state $\ket{\uparrow'}$ to the ground-state qubit level $\ket{\downarrow}$, $\gamma_2$ is the optical pure dephasing rate in the bulk, and $\Delta_\omega$ is the detuning between optical transitions of two Yb ions. Here, we assume the decay rate from $\ket{\uparrow'}\mapsto\ket{\uparrow}$ is negligible. Indeed, as we argue later, this is the case for Yb:YVO. We estimate the optical pure dephasing rate with the relation
$\gamma_2 = 1/T_{2o} - \gamma_1/2 \approx 9.12$ kHz, where $T_{2o}=91\mu$s is the optical coherence time of Yb ions in YVO crystal bulk \cite{kindem2018characterization}. This results in a fidelity of $\mathrm{F_{PI}} = 0.51$.

This reduced fidelity is attributed to the Yb ion optical pure dephasing rate captured by $T_{2o}$, which is influenced by both spin decoherence and spectral fluctuations of the optical transition that occur faster than the optical lifetime. However, knowing that the ground state spin coherence time $31$ ms \cite{kindem2020control} is already two orders of magnitude larger than the unenhanced optical lifetime $T_{1o}=267\mu\text{s}$ \cite{kindem2018characterization}, the contribution of the ground state spin dephasing to the optical decoherence is negligible. Furthermore, any amount of Purcell enhancement needed to overcome the optical dephasing rate will further reduce the impact of the ground state spin decoherence and hence it can be safely neglected in the analysis.

While this gate scheme does not inherently require a cavity to be performed, below, we discuss how employing a cavity can enhance the fidelity of the gate between two Yb ions. In case of using a cavity \cref{f_bk} is modified to
\begin{equation}\label{f_bk_cavity}
\mathrm{F'_{PI}} = \frac{1}{2}\left(1 + \frac{\gamma_1 ^{'2}}{(\gamma_1'+2\gamma^{*})^2+\Delta_\omega^2}\right),
\end{equation}
where $\gamma_1'=\gamma_{1r}F_p+\gamma_1$ defines the Purcell-enhanced optical decay rate of the ion in the presence of the cavity, $\gamma_1=\gamma_{1r}+\gamma_{1nr}$ is Yb ion decay rate with radiative ($\gamma_{1r}$) and non-radiative decay rate ($\gamma_{1nr}$) parts, and $\gamma^{\ast}$ is the optical pure dephasing rate in the cavity.
The Purcell factor in both strong coupling and bad-cavity regimes is defined as \cite{wein2021modelling} 
 \begin{equation}\label{purcell}
     F_p=\frac{R\kappa}{\gamma_r(\kappa+R)},
 \end{equation}
with $\kappa$ being the cavity decay rate, and 
\begin{equation}
 R=\frac{4g^2(\kappa+\gamma_1+2\gamma^{*})}{(\kappa+\gamma_1+2\gamma^{*})^2+4\Delta_\omega^2}.
\end{equation}

\begin{figure}[t]
\centering
\includegraphics[width=0.5\textwidth]{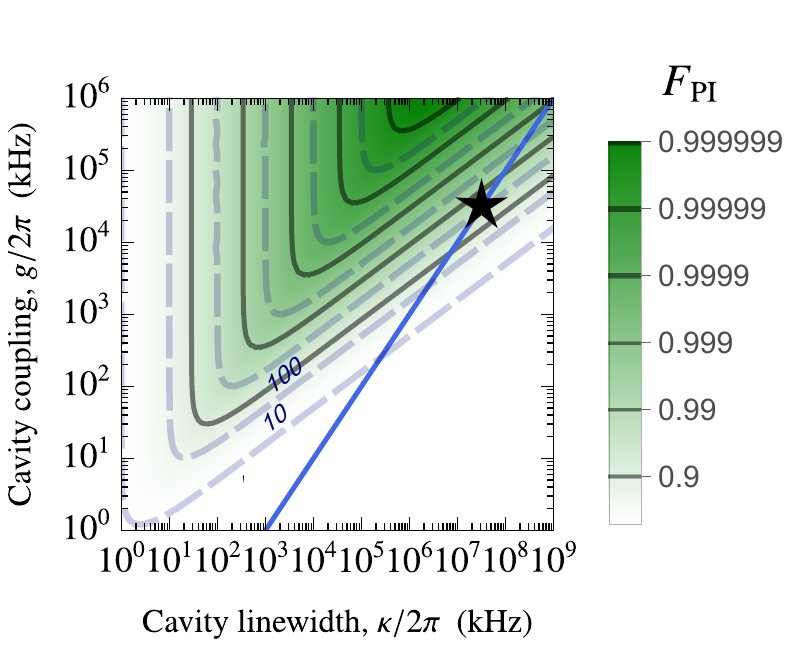}
\caption{Fidelity of photon interference scheme as a function of cavity parameters. 
Here we use a slightly larger pure dephasing rate of $\gamma^{*}=23.63$ kHz compared to bulk, which is expected for ions in a cavity \cite{kindem2020control} and is consistent with the dephasing rate we considered for the photon scattering scheme (see \cref{ps implemnt} for more details.).
The parameters for this figure are identical to those of the photon scattering–based gate, detailed in \cref{ps implemnt} and summarized in \cref{tab:YbYVO_parameters}. We also assume that spectral diffusion is mitigated, so the optical transition frequencies of the Yb ions are in perfect resonance with each other, i.e., $\Delta_\omega = 0$ (see \cref{impl:MD} and \cref{ps implemnt} for more information).
We also note that dark counts are not included in our gate modeling, since we assume (as
discussed in \cref{dark}) moderate efficiencies are achievable.
With these assumptions, the current experimental parameters reach a fidelity of 0.95, which is indicated by the star symbol. The blue line is studied separately in \cref{comparision3}. }
\label{imp:PI}
\end{figure}

We demonstrate how the fidelity of this scheme varies with different cavity parameters in \cref{imp:PI}. Additionally, we determine the necessary Purcell factor for achieving a specific fidelity threshold.

Assuming resonant excitation results in photons with a Lorentzian spectral shape, and since this scheme takes two emissions, we consider the gate time for this scheme twice the gate time definition in the photon scattering gate $T_{g,PI}=14T_{1p}^\prime$. Therefore, we allow at most $7T_{1p}$ for each emission time bin. However, it should be noted that here $T_{1p}^\prime$ denotes the Purcell-enhanced lifetime of the photon set by the emitter, which is fixed by $T_{1o,\mathrm{cav}}$. In contrast, in the photon scattering scheme, the photon originates from an external source and not the ion. Furthermore, this gate time neglects the time taken to perform the necessary spin-flip operation between the emission events. However, as discussed in \cref{impl:MD}
we assume that single-qubit operations can be implemented much faster than the time it takes to emit two photons.\\

In the simulations of this gate scheme, we use the ground state spin coherence time of $T_{2s,g,\mathrm{cav}}=31$ ms measured using Carr-Purcell Meiboom-Gill (CPMG) decoupling sequences \cite{kindem2020control}. Regarding the compatibility of using dynamical decoupling (DD) techniques to enhance the coherence time with the photon interference protocol, we point out that the protocol itself already requires at least one microwave $\pi$ pulse between the two photon-emission rounds to perform the spin bit-flip. 
One can, however, extend this by applying additional microwave $\pi$ pulses. 
The current experiment on erbium spin-photon entanglement \cite{uysal2025spin} uses more advanced sequences such as XYn, where the entire sequence happens between the two-round photon emissions.
It is important to note that the total number of pulses must be odd; otherwise, if an even number of DD pulses are applied, they would cancel the net bit-flip effect, violating the protocol's requirement for a spin inversion between photon emissions.
To avoid overlap between the pulses and the photon emission process, they apply the DD pulses not only to the ground state but also to the excited state, thereby ensuring that photons are emitted only when there has been an odd number of bit-flips in the ground state~\cite{uysal2025spin}. This effectively shelves the excited state, and the remaining photons are emitted after the second DD pulse.  
This approach may be adaptable to the Yb:YVO system.  However, a more in-depth investigation of such an implementation goes beyond the scope of the present paper.

We note that in this analysis, similar to the other photon-detection-based scheme discussed in the previous section (e.g., photon scattering-based scheme), we did not explicitly include the effects of detector dark counts or noise counts due to the photons emitted from nearby ions on the fidelity. The reason is that, unlike the intrinsic properties of the system, dark and noise counts are not considered a fundamental limitation for photon detection–based gate protocols. In fact, technological progress aims to both enhance single-photon collection and detection efficiencies and reduce dark and noise count rates. 
In \cref{dark}, we demonstrate that for the photon interference-based scheme, dark counts are not the primary limitation at moderately good efficiencies, which underscores the importance of improving the overall efficiency. Enhancing the efficiency directly increases the signal-to-noise ratio by raising the probability of detecting good counts.
The overall efficiency is constrained by multiple sources of loss that occur throughout the process, from photon emission to detector registration. Mitigating these losses can lead to a substantial improvement in the total efficiency.    
Moreover, the noise count problem can be reduced by insulating the experiment from the environmental light.
A similar calculation can be done for the photon scattering-based scheme.
Therefore, we did not incorporate dark-count effects for the photon detection-based schemes in our roadmap study, as our goal was to present an idealized scenario that highlights which intrinsic atomic properties fundamentally limit gate performance---even in the case where future technological advances (e.g., higher collection efficiencies, better cavity fabrication, or detectors with negligible dark count rates) address current technical imperfections. 
Nevertheless, we point out that dark counts and noise counts are currently one of the limiting factors for fidelity in both two-ion entanglement and spin–photon entanglement with rare-earth ions \cite{uysal2025spin,ruskuc2025multiplexed}.
Although we have not included this effect in our simulations, its impact on the fidelity of photon interference–based schemes has been analyzed in \cite{wein2020analyzing}.

In this scheme, as in the photon scattering scheme, qubit levels are defined using two of the lowest hyperfine levels in the ground state, along with an ancillary level in the excited state. The zero field hyperfine energy-level diagram and allowed transitions for the Yb:YVO system are given in Fig. 1 of Ref. \cite{kindem2018characterization}. Based on this level structure, for a photon-interference–based scheme, a reasonable choice is to define the qubit states (i.e., $\ket{\uparrow},\ket{\downarrow}$) using the hyperfine levels $\ket{3}_g$ and $\ket{4}_g$, and to use $\ket{1}_e$ as the auxiliary level (i.e., $\ket{\uparrow'}$). As shown in this diagram, the transition $\ket{3}_g \leftrightarrow \ket{1}_e$ is symmetry-forbidden, while transition $\ket{4}_g \leftrightarrow\ket{1}_e$ is allowed with a branching ration of $\beta_{\ket{4}_g \leftrightarrow\ket{1}_e}\approx0.35$. However, applying a Purcell enhancement of $\sim 335$ has modified the branching ratio of this transition to  $\beta^{cav}_{\ket{4}_g \leftrightarrow\ket{1}_e}=0.9968$ \cite{kindem2020control}.

Regarding decay to higher ground state CF levels, for the Yb:YVO system, the splitting between the CF levels is on the order of THz \cite{kindem2018characterization}. Consequently, photons emitted to those higher ground state CF levels have frequencies significantly different from those emitted to the lowest ground state CF level ($^2F_{7/2}(0)$), and hence, they can be filtered out.

To avoid off-resonance excitations, similar to other schemes, the Rabi frequency of the excitation pulse is chosen to be much smaller than the qubit-level splitting (0.675 GHz \cite{kindem2018characterization}). For example, similar to the magnetic dipolar gate, a Rabi frequency of 10 MHz can be used. This corresponds to a pulse duration of $0.31~\mu\text{s}$, which is shorter than the Purcell-reduced lifetime of $2.27~\mu\text{s}$ \cite{kindem2020control} and makes the single-qubit gate operations possible.

Similarly to the previous subsections, we neglect the effect of spectral diffusion during the short gate operation time in this scheme.

\section{Two-qubit Gate scheme Comparison}\label{comparing}

In our comparison of gate schemes, we focus on the two-qubit interaction, assuming that within each scheme the single-qubit gate errors are negligible. Nevertheless, we recall that the magnetic dipolar gate requires two additional single-qubit Z gates, and the photon interference scheme requires a measurement basis change to achieve a CZ gate, whereas the photon scattering gate implements it directly.

We now analyze all three 
two-qubit gate schemes, among which the magnetic dipolar gate is fully deterministic, the photon scattering gate is near deterministic, and the photon interference-based scheme is a probabilistic scheme with a maximum success probability of $50\%$.

We begin our comparison with a number of qualitative remarks.
In terms of experimentally controlling the energy levels, the magnetic dipolar gate requires two energy levels in both ground and excited states while photon scattering and photon interference-based can be performed utilizing only two energy levels in the ground and one auxiliary level in the excited state.
Among the gates, since dipolar-type gates require direct interactions between the qubits (ions), they are highly limited by the spatial separation between the individual ions and hence, the effectiveness of the magnetic dipolar gate falls off significantly if the distance exceeds a few nanometers. At such small distances, the spatial addressability of individual ions is a challenge. Conversely, 
identifying a pair of ions within the cavity is relatively straightforward, since the ions can be separated by distances on the order of cavity lengths. However, achieving precise addressability of individual ions, particularly in the microwave domain, remains challenging even with the presence of a cavity.
Moreover, designing such gates can be challenging due to the difficulty in fabricating nanophotonic crystal cavities, which are limited by the availability of high-quality thin films.
Additionally, bringing optical transitions into resonance-- with each other and the cavity mode in photon scattering schemes-- requires more advanced experimental techniques.
The photon interference-based scheme offers greater flexibility in terms of separation distance, as it does not rely on direct qubit-qubit interactions.
For instance, a cavityless version of the photon interference-based gate scheme has been successfully implemented between two NV centers separated by 3 meters \cite{bernien2013heralded}. While the photon interference-based gate can also be operated without a cavity, incorporating a cavity for rare-earth ions significantly improves the fidelity through Purcell enhancement.
Nevertheless, even in the enhanced case, this scheme requires a delay line to recombine the time bins and implement a Hadamard gate. This necessitates adding an active switch and several kilometres of optical fibre prior to detection, which incurs losses and increases the experimental cost. However, with the use of hollow-core fibres (HCFs) \cite{petrovich2025broadband}, high transmission rates at Yb wavelengths can be expected in the future, at the expense of additional experimental complexity.
An alternative solution could be utilizing quantum memories. This practical challenge underscores the need to design cavities with higher cooperativity.

We also highlight the challenge of dark and noise counts in the photon-detection-based schemes. However, as our calculations in \cref{dark} conclude, for moderately good efficiencies, dark counts are not expected to be a major issue, emphasizing the importance of enhancing overall efficiency. As a potential mitigation strategy, we highlight that the current total single-photon detection efficiency of approximately $\sim 1\%$ can be increased, for example, by enhancing the photon collection efficiency through improvements in the Purcell factor and the cavity--waveguide coupling $(\kappa_{\text{in}}/\kappa)$. Currently, the primary limitations on detection efficiency arise from the waveguide--cavity coupling $\kappa_{\text{in}}/\kappa \approx 0.12$ and the waveguide--fiber coupling $\approx 0.24$~\cite{ruskuc2025multiplexed, kindem2020control}. 
Enhancing these parameters would increase both the detection efficiency and the entanglement generation rate. In addition, to mitigate noise counts,~\cite{ruskuc2025multiplexed} suggests reducing resonant noise counts by using smaller Rabi frequencies, though this may increase pulse errors. Another approach is to reduce the concentration of Yb ions in the crystal.

\begin{figure}[t]
    \centering
    \includegraphics[width=1\linewidth]{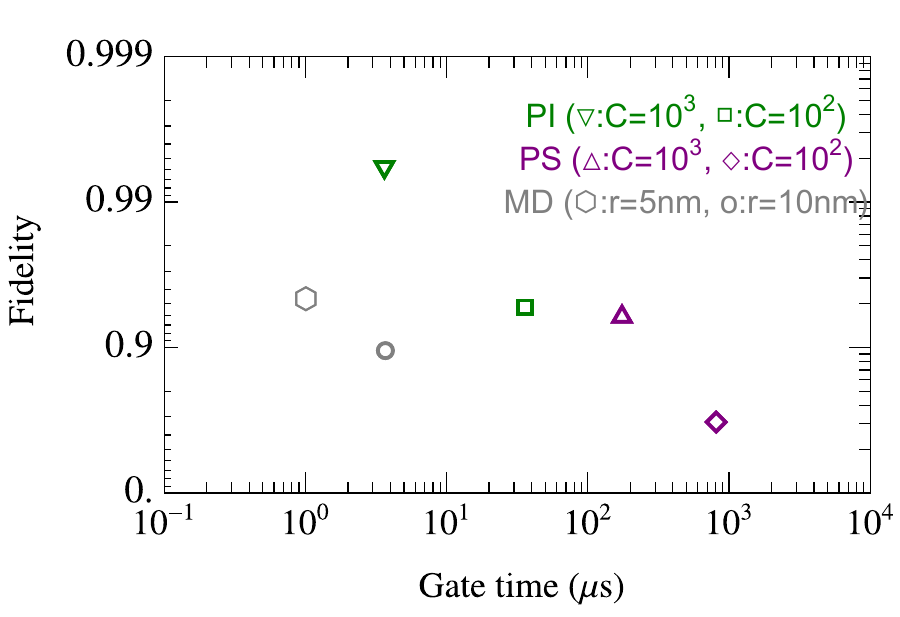}
    \caption{The fidelities and corresponding gate times of the gate schemes discussed in this work. Here, we present two achievable values for the fidelity and gate time of each scheme, based on experimentally attainable parameters for rare-earth ions. This includes varying the ion separation distance ($r$) for the magnetic dipolar gate and the cavity cooperativity ($C$) for the other schemes.}
    \label{fig:gate time}
\end{figure}

\begin{figure}[t]
\centering
\includegraphics[width=0.5\textwidth]{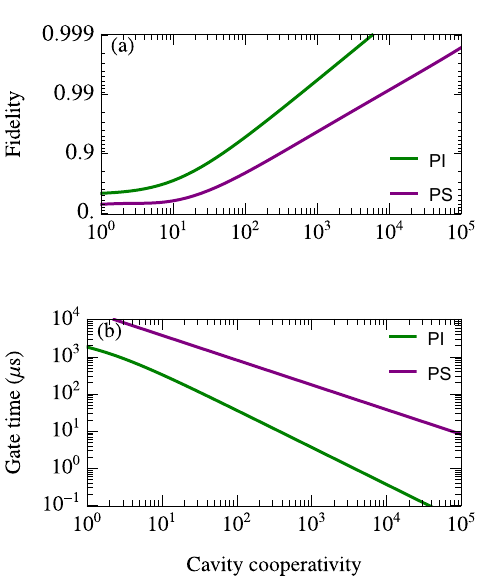}
\caption{(a) Fidelity and (b) gate time of the cavity-based gate schemes as a function of cavity cooperativity for a slice of the 2D plots corresponding to $g/\kappa \simeq 0.001 $ that intersects with the star marks in \cref{imp:PS} and \cref{imp:PI}.}
\label{comparision3}
\end{figure}

For a quantitative comparison of the gates, we present the fidelities and associated gate times in \cref{fig:gate time}. The symbols in \cref{fig:gate time} indicate that within a specific gate scheme, a shorter gate time is associated with higher fidelity. 
We point out that for gate time calculations, we focus on the inherent gate time of the schemes. Specifically for the non-deterministic gate schemes, we separate the success probabilities from the inherent gate time. Hence, we focus on the gate time, which is limited by the photon bandwidth as it reflects the device response required to reach a target fidelity, rather than the average rate at which gates can be executed once the exact protocol is defined and implemented.
Examining the fidelity equations for each scheme, we find that the most significant technology-related degrees of freedom are the ion separation for the magnetic dipolar-based interaction scheme and the cavity cooperativity for the other schemes.
Therefore, we analyze gate times and fidelity scalings with respect to technology-related parameters, illustrating how advances in these parameters can improve both fidelities and gate times.
Achieving both shorter gate time and higher fidelity requires either smaller ion separation for the magnetic dipolar scheme or higher cooperativity for the photon scattering and photon interference-based 
schemes which pose practical restrictions. Nonetheless, the specific cooperativity requirements are scheme-dependent. A comparison between different schemes demonstrates that the photon interference-based scheme achieves superior fidelity and shorter gate times than the photon scattering scheme across both lower (e.g., $C = 100$) and higher (e.g., $C = 1000$) cooperativity regimes.
The role of cooperativity in determining the performance of different cavity-based schemes is examined in more detail in \cref{comparision3}.

Also, for the distances considered here, the magnetic dipolar gate achieves shorter gate times than the other two schemes. However, attaining high fidelities remains challenging compared to the photon interference-based scheme at cooperativity of $C=1000$.

The limiting parameter for cavity-based gate schemes is currently achievable cavity cooperativity. Therefore, in \cref{comparision3} we compare the performance of the cavity-based gate schemes in terms of required cooperativity. The photon interference-based scheme is more applicable than the photon scattering scheme with the currently reported cavity cooperativity for Yb
to implement a two-qubit gate. 
To summarize the cooperativity dependence of the different gate schemes, \cref{comparision3} shows that the infidelity of the photon interference-based gate scales as $1/C$, while for the photon scattering
scheme, the infidelity scales as $1/C^{2/3}$.

\section{Conclusion and Outlook}\label{disc}

In this paper, we discussed three particularly attractive two-qubit gate schemes—arguably the most promising choices—given the current state of the art of the technology of the Yb:YVO system, including a deterministic approach, namely magnetic dipolar gate, a near deterministic approach, namely photon scattering scheme and a probabilistic scheme based on photon interference.
We introduced a new approach for calculating the state and gate fidelities of a desired gate scheme. With a focus on state fidelity for quantum communication use cases, we examined their feasibility for implementation and compared their performance based on state-of-the-art technology.

Our analysis showed that each gate scheme has its advantages and limitations. The photon interference-based scheme outperforms the photon scattering gate by offering higher fidelity and shorter gate times as cooperativity increases. However, the photon interference-based scheme operates probabilistically, whereas the photon scattering scheme is nearly deterministic. In contrast, the magnetic dipolar gate performs independently of a cavity, making it a promising alternative. However, it relies on tightly localized ions to achieve fast and high-fidelity operations.\\

We note that rare-earth systems currently have collection efficiencies much less than one. Here, our analysis focuses on fidelity estimates and physical limitations on gate quality, with a detailed efficiency analysis left outside the scope of this work, as fidelity and efficiency are, to first order, independent considerations. Nevertheless, we acknowledge that very low efficiencies would extend the relevant time scales for the non-local gates (photon interference-based and photon scattering) if they are repeated until success. For example, we have discussed the effect of efficiency on the dark count problem in \cref{dark}. In addition, propagation time should also be considered based on the application and layout of the emitters, thereby making practical effects such as spectral diffusion and limited coherence times increasingly significant.

To provide a broader perspective, we note that the schemes considered here are not the only available options, and other approaches may also be viable.
We also highlight other potential gate implementation techniques
that experimentalists might find valuable, depending on their experimental facilities
and specific needs.
For example, an alternative approach for implementing a deterministic two-qubit gate involves virtual photon exchange within a cavity \cite{asadi2020cavity}. This scheme has been widely used as a readout protocol in superconducting qubits \cite{majer2007coupling,blais2004cavity}.
Applying it to rare-earth ions requires a high-cooperativity cavity \cite{asadi2020protocols}.
In \cref{fid:virtual}, we re-calculate the fidelity of this scheme. This example illustrates a scenario involving a non-Hermitian error Hamiltonian, which we analyze using the perturbative method developed in this work. With current technology and in the absence of excited state spin dephasing, this scheme achieves a fidelity of $50\%$.
However, as demonstrated in \cite{asadi2020protocols, wein2021modelling}, monitoring the cavity emission can enhance the fidelity of this gate scheme through the post-selection of successful gates, though this comes at the cost of rendering the scheme non-deterministic.
Another scheme of interest for optically long-lived rare-earth ions is introduced in \cite{martin2019single}. They demonstrated that the photon interference-based scheme discussed earlier can be modified to become nearly deterministic by incorporating feedback during the measurement process.
Exploring additional photon-mediated schemes such as those discussed in \cite{beukers2024remote} and carefully analyzing their implementation feasibility for Yb:YVO would be an interesting direction for future research.

Looking forward, one can leverage the optimal gate protocol established in this work for designing a quantum repeater between individual Yb ions and also for distributed quantum computing tasks.
While this work focuses on YVO, it is worth noting that similar approaches presented here could also be applied to other crystals, such as YSO, if a nanophotonic cavity can be designed for $^{171}\text{Yb}^{3+}$:YSO to assess the performance of cavity-based gate schemes.
Although our comparison study targets $^{171}\text{Yb}^{3+}$:YVO system, similar gate comparisons could also be done for other systems (other RE ions in solids, other kinds of defect centers etc.), where the most favourable gates will likely depend on the system parameters such as radiative lifetimes and coherence times as well as the experimental state of the art, e.g. for defect localization and cavity cooperativity.

\section*{Acknowledgement}
This work is funded by the NSERC Alliance quantum consortia grants ARAQNE and QUINT and the NRC High-throughput Secure Networks (HTSN) challenge program.
We thank Andrei Faraon,
Sophie Hermans,
Mikael Afzelius, Arsalan Motamedi, Jiawei Ji, Nasser Gohari Kamel and Andrei Ruskuc for useful discussions. We would also like to thank Geza Giedke and anonymous reviewers for helpful comments.

\bibliographystyle{unsrt}
\bibliography{ref}

\clearpage

\section*{Supplementary Material}

\setcounter{section}{0}
\renewcommand{\thesection}{S\Roman{section}}

\setcounter{equation}{0}
\setcounter{figure}{0}
\setcounter{table}{0}

\renewcommand{\theequation}{S\arabic{equation}}
\renewcommand{\thefigure}{S\arabic{figure}}
\renewcommand{\thetable}{S\arabic{table}}

\section{Perturbative fidelity computation}\label{app:perturbation}
In this section, we elaborate on our techniques presented in \cref{sec:state-fid}. We show how one can compute the infidelity up to the desired order, and in particular, we provide a derivation for \cref{eps-first-order}, \cref{eq:non-hermit-expression}, and \cref{eps-second-order}.\\

We rewrite \cref{eq:lindbladian} as
\begin{align}\label{eq:app-lindbladian-fid}
\dot{\rho} = -\frac{i}{\hbar}[H_g+\delta \tilde H_e,\rho] + \mathcal D[\rho], 
\end{align}
with $\mathcal D[\rho] = \sum_{k}\gamma_k L_k \rho L_k^\dagger -  \frac{\gamma_k}2 \{L_k^\dagger L_k,\rho \}$. More specifically, we may write $\mathcal D_k$ as the superoperator considering the effects of the $k$-th term, i.e., $\mathcal D_k[\bullet] := L_k \bullet L_k^\dagger - \frac12 \{L_k^\dagger L_k, \bullet\}$.
We then begin with considering evolution in the interaction picture. Let
\begin{align}\label{eq:interaction-def}
\begin{split}
\rho_I(t) &= U^\dagger(t) \rho(t) U(t)\\
\mathcal D_I[\bullet] &= U^\dagger(t) \mathcal D[U(t) \bullet U^\dagger(t)] U(t)\\
H_{e,I} &= U^\dagger(t) \tilde H_e U(t), 
\end{split}
\end{align}
where the unitary operator $U(t)$ is defined as $U(t) = \mathscr{T}\exp\left(-\frac{i}{\hbar}\int_{0}^{T_g}H(t') \mathrm{d}t'\right)$, and the subscript $I$ denotes quantities in the interaction picture. Moreover, we note the operator $\mathcal D_{k,I}$ can be written via $L_{k,I} = U(t) L_k U(t)^\dagger$ as
\begin{align}\label{eq:lind-interaction}
\mathcal D_I[\bullet]  = \sum_{k} L_{k,I} \bullet L_{k,I}^\dagger - \frac12 \{L_{k,I}^\dagger L_{k,I},\bullet \}.
\end{align}
Differentiating $\rho_I$ with respect to time gives
\begin{align}\label{eq:interaction-evolution}
\begin{split}
\frac{\mathrm d}{\mathrm dt} \rho_I(t) &= U^\dagger(t) \frac{i}{\hbar}[H_g,\rho(t)]U(t) + U(t)^\dagger \dot{\rho}(t) U(t)\\
&\overset{(i)}{=} U(t)^\dagger\left( -\frac{i}{\hbar}\delta[H_e,\rho(t)]+ \mathcal D[\rho(t)]\right) U(t)\\
&\overset{(ii)}{=} -\frac{i}{\hbar}\delta[H_{e,I}(t),\rho_I(t)] + \mathcal D_I [\rho_I(t)],
\end{split}
\end{align}
where (i) follows from \cref{eq:app-lindbladian-fid} and (ii) is due to \cref{eq:interaction-def}. Furthermore, a simple calculation demonstrates that, employing the definition we provided in \cref{eq:fidelity-definition}, we obtain
\begin{align}\label{eq:fidelity-interaction}
F = \langle \psi(0) | \rho_I(T_g) | \psi(0) \rangle.
\end{align}
In order to compute $\rho_I(T_g)$, we note that one can obtain $\rho_I$ at any time, say $t$, by rewriting \cref{eq:interaction-evolution} in the integral form \footnote{Note that we are implicitly using the fact that $\rho_I(0)=\rho(0)=\ket{\psi(0)}\bra{\psi(0)}$.}
\begin{align}
\begin{split}
\rho_I(t) &= \rho(0) -\frac{i\delta}{\hbar} \int_0^{t} [H_{e,I}(t'),\rho_I(t')] \, \mathrm dt' \\ &\qquad+ \int_{0}^{t} \mathcal D_{I}(t')[\rho_I(t')]\, \mathrm dt'.
\end{split}
\end{align}
Using this equation recursively, we get the following first and second-order approximate solution
\begin{align}\label{eq:1st-rho}
\begin{split}
\rho^{(1)}_I(T_g) &= \rho(0)-\frac{i\delta}{\hbar} \int_0^{T_g} [H_{e,I}(t'),\rho(0)] \, \mathrm dt' \\ &\qquad+ \int_{0}^{T_g} \mathcal D_{I}(t')[\rho(0)]\, \mathrm dt' \;\; (1^{\text{st}}\text{ order}),\\
\end{split}
\end{align}
and
\begin{align}\label{eq:2nd-rho}
\begin{split}
\rho^{(2)}_I(T_g) &= \rho(0) -\frac{i\delta}{\hbar} \int_0^{T_g} [H_{e,I}(t'),\rho^{(1)}(t)] \, \mathrm dt' \\ &\qquad+ \int_{0}^{T_g} \mathcal D_{I}(t')[\rho^{(1)}(t)]\, \mathrm dt' \; \;(2^{\text{nd}}\text{ order}).\\
\end{split}
\end{align}
More generally, one can recursively employ $\rho_I^{(j)}(T_g) = \frac{i\delta}{\hbar} \int_0^{T_g} [H_{e,I}(t'),\rho^{(j-1)}(t)] \, \mathrm dt' + \int_{0}^{T_g} \mathcal D_{I}(t')[\rho^{(j-1)}(t)]\, \mathrm dt'$ to obtain the $j$-th order approximation to $\rho_I(t)$. We note that the difference between $\rho_I(t)$ and $\rho_I^{(j)}$ is of the order of $(|\delta|+\sum_k |\gamma_k|)^{j+1}$ i.e., $\norm{\rho_I(t) - \rho^{(j)}_I(t)} = O(|\delta|+\sum_k |\gamma_k|)^{j+1}$. 
Using such expansions, together with \cref{eq:fidelity-interaction} we get
\begin{align}\label{eq:pert-fid}
\begin{split}
F = &1-\langle \psi(0)| \rho_I^{(j)}|\psi(0)\rangle\\
&+ 
O\left( T_g^{j+1}\left(\frac{\norm{\tilde H_e}}{\hbar}|\delta|+\sum_k |\gamma_k|\right)^{j+1}\right).
\end{split}
\end{align}
In what follows, we elaborate on the computation of \cref{eq:pert-fid} and demonstrate calculations that yield to \cref{eps-first-order}. Next, in \cref{sec:app-non-hermitian-derivation}, we consider the non-Hermitian perturbation to the Hamiltonian and derive \cref{eq:non-hermit-expression}.

\subsection{Derivation of \cref{eps-first-order} and \cref{eps-second-order}}\label{sec:app-derivation}
Let us start by examining the fidelity for the first-order approximation $\rho_I^{(1)}$ in \cref{eq:1st-rho}. This gives
\begin{align}
\begin{split}
&1-F =\\
&-\frac{i\delta}{\hbar} \int_{0}^{T_g} \langle \psi(0)| \, [H_{e,I}(t'),\ket{\psi(0)}\bra{\psi(0)}] \, \ket{\psi(0)} \\
&\qquad + \int_{0}^{T_g}\bra{\psi(0)}\mathcal D_{I}\left[\ket{\psi(0)}\bra{\psi(0)}\right]\ket{\psi(0)} \\
&\qquad+ 
O\left( T_g^2\left(\frac{\norm{\tilde H_e}}{\hbar}|\delta| + \sum_{k} |\gamma_k|\right)^2 \right).
\end{split}
\end{align}
Note that $\langle \psi(0)| [H_{e,I}(t'),\ket{\psi(0)}\bra{\psi(0)}] \ket{\psi(0)}=0$, which means that the first-order contribution of $\delta$ (i.e., reversible errors) to infidelity is zero. However, accounting for first-order irreversible errors, we have
\begin{align}\label{eq:sand-2}
\begin{split}
&\bra{\psi(0)}\mathcal D_{I}\left[\ket{\psi(0)}\bra{\psi(0)}\right]\ket{\psi(0)} = \\
&\qquad\qquad \sum_{k} \gamma_k \langle \psi(0)|L_{k,I}|\psi(0)\rangle \langle \psi(0)|L^\dagger_{k,I}|\psi(0)\rangle\\
&\qquad\qquad -\frac12 \bra{\psi(0)} \{L_{k,I}^\dagger L_{k,I}, \ket{\psi(0)}\bra{\psi(0)}\} \ket{\psi(0)},
\end{split}
\end{align}
which follows directly from \cref{eq:lind-interaction}. Note that 
\begin{align}\label{eq:sand-3}
\begin{split}
\langle\psi(0)|L_{k,I}\ket{\psi(0)} &= \bra{\psi(0)} U^\dagger(t) L_k U(t)\ket{\psi(0)} \\
&= \langle L_k(t)\rangle,
\end{split}
\end{align}
as we are adapting the notation $\langle \mathcal O(t)\rangle:=\bra{\psi(t)} \mathcal O\ket{\psi(t)}$. Using a similar argument for the second term on the right-hand side of \cref{eq:sand-2} we get
\begin{align}
1-F = \epsilon_L^{(1)} + O\left( T_g^2\left(\frac{\norm{\tilde H_e}}{\hbar}|\delta| + \sum_{k} |\gamma_k|\right)^2 \right),
\end{align}
with
\[
\epsilon_L^{(1)} = \sum_{k} \int_{0}^{T_g} \gamma_k \left( \langle L_k^\dagger(t') L_k(t') \rangle - |\langle L_k(t') \rangle|^2 \right) \, \mathrm dt',
\]
which is identical to \cref{eps-first-order}. We proceed to the second term correction by computing $\bra{\psi(0)}\rho^{(2)}_I - \rho^{(1)}_I\ket{\psi(0)}$. Note that $ \epsilon_L^{(1)} + \bra{\psi(0)}\rho^{(2)}_I - \rho^{(1)}_I\ket{\psi(0)}$ gives us the second order approximation to $F$. However, as we are merely interested in the first non-zero perturbation terms, we will keep only the terms including $\delta^2$ and $\delta\cdot 
\gamma_k$ as the first-order contribution of $\delta$ is zero.
We have
\begin{align}
\begin{split}
&\rho^{(2)}_I - \rho^{(1)}_I =\\
&-\frac{\delta^2}{\hbar^2} \int_0^{T_g}\int_{0}^t[H_{e,I}(t'),[H_{e,I}(t''),\rho_0]]\, \mathrm dt' \, \mathrm dt''\\
&-\frac{i\epsilon}{\hbar}\bigg(\int_0^{T_g}\int_0^{t} [H_{e,I}(t'),\mathcal D_I(t'')[\rho(0)]] \, \mathrm dt'\, \mathrm dt''\\
&- \int_0^{T_g}\int_0^{t} \mathcal D_I(t')[[H_{e,I}(t'),\rho_0]] \, \mathrm dt'\, \mathrm dt''\bigg)\\ &+O(T_g^2(\sum_k |\gamma_k|)^2)\\
&+ O\left( T_g^3\left(\frac{\norm{\tilde H_e}}{\hbar}|\delta|
+ \sum_{k} |\gamma_k|\right)^3 \right).
\end{split}
\end{align}
By a straightforward calculation, we get
\begin{align}
\begin{split}
&\bra{\psi_0}[H_{e,I}(t'),[H_{e,I}(t''),\rho_0]]\ket{\psi_0} =\\
&\qquad \Re\bigg(\langle H_{e}(t') H_{e}(t'')\rangle\\
&\qquad -\langle H_{e}(t')\rangle\langle H_e(t'')\rangle \rangle\bigg), 
\end{split}
\end{align}
which corresponds to
$\epsilon^{(2)}_{HH}$ as in \cref{eps-second-order}. Similarly, we get
\begin{align}
\begin{split}
&-i\bra{\psi_0}[H_{e,I}(t'),\mathcal D_I[\rho_0]] \ket{\psi_0}= \\ 
&\qquad \Im\bigg(\sum_{k} \gamma_k \langle H_{e}(t') L_k(t'')\rangle \langle L_k(t'')\rangle\\
&\qquad -2\langle H_e(t') L_k(t'')\rangle \langle L_k^\dagger(t'')\rangle\bigg),
\end{split}
\end{align}
and
\begin{align}
\begin{split}
&-i\bra{\psi_0} \mathcal D_I(t')[[H_{e,I}(t'),\rho_0]] \ket{\psi_0}=\\
&\qquad\Im\bigg( \langle L_k^\dagger (t') L_k(t') H_e(t'')\rangle\\
&\qquad -2\langle L_k(t')H_e(t'')\rangle \langle L_k^\dagger (t')\rangle\bigg),
\end{split}
\end{align}
which gives
\begin{align}
\begin{split}
\epsilon_{LH}^{(2)} &= \sum_{k} \frac{\gamma_k\delta}\hbar \int_{t'=0}^{T_g} \int_{t''=0}^{t'} \mathrm{Im} \bigg( \langle H_{e} (t')L_k^\dagger(t'') L_k(t'') \rangle \\
&\qquad \qquad- 2 \langle H_{e}(t') L_k(t'') \rangle \langle L_k^\dagger (t'') \rangle\bigg) \, \mathrm dt'' \, \mathrm dt' \\
&+ \sum_{k} \frac{\gamma_k\delta}\hbar \int_{t'=0}^{T_g} \int_{t''=0}^{t'} \mathrm{Im}\bigg( \langle L_k^\dagger(t')L_k(t') H_e(t'') \rangle  \\
&\quad  - 2 \langle L_k(t') H_e(t'') \rangle \langle L_k^\dagger(t') \rangle \bigg) \, \mathrm dt'' \, \mathrm dt'.
\end{split}
\end{align}

\subsection{Derivation of \cref{eq:non-hermit-expression}}\label{sec:app-non-hermitian-derivation}

We note that for a non-Hermitian perturbation, the evolution can be written as
\begin{align}
\dot{\rho} = -\frac{i}{\hbar}\left([H_g,\rho] + \delta(\tilde H_e \rho - \rho {\tilde H_e}^\dagger)\right) + \mathcal D[\rho],
\end{align}
where $\tilde H_e$ is the non-Hermitian perturbation. Employing the approach introduced earlier to get \cref{eq:1st-rho}, we find that
\begin{align}
\begin{split}
\rho_I^{(1)}(T_g) = \rho(0) &- \frac{i\delta}{\hbar} \int_{0}^{T_g} \left(  H_{e,I}(t')\rho_0 - \rho_0{ H_{e,I}}^\dagger(t') \right)\\
&+\int_{0}^{T_g} \mathcal D_{I}(t')[\rho(0)]\, \mathrm dt'.
\end{split}
\end{align}
A straightforward calculation yields
\begin{align*}
\bra{\psi(0)}\! \left(\!  H_{e,I}(t')\rho_0\! -\!\rho_0{H_{e,I}}^\dagger(t')\! \right)\! \ket{\psi(0)} =
&\langle \tilde H_e(t') \rangle \\
&- \langle {\tilde H_e}^\dagger(t')\rangle.
\end{align*}
This gives us the $\epsilon_H^{(1)}$ as in \cref{eq:non-hermit-expression}.

\section{Average gate fidelity}\label{gate-fids}

In this section, we demonstrate how the state fidelity computation obtained through the perturbative approach can be extended to compute the average gate fidelity. To this end, we use the entanglement fidelity, which is closely connected to the average gate fidelity.

Let $\mathcal E[\bullet] = \mathscr{T}\exp(\int_{0}^{T_g} \mathcal L_I(t') \mathrm dt')[\bullet]$ be the channel representing the evolution under the interaction picture's Lindbladian. So far, we have considered the computation of 
\begin{align}
F_{\psi_0} = \langle \psi(0)| \mathcal E[\psi(0)]|\psi(0)\rangle
\end{align}
which is the fidelity of between the output state of the Lindbladian evolution and the ideal state of the computation, starting from $\ket\psi(0)$. However, in quantum computation, one is often interested in `gate fidelity.' Important instances of gate fidelities include the minimum, average, and entanglement fidelity \cite{nielsen2001quantum}. These three notions, in our case, can be formulated as
\begin{align}\label{eq:gate-fids}
\begin{cases}
F_{\min} = \min_\psi \langle \psi|\mathcal E[\psi]|\psi\rangle\\
F_{\mathrm{avg}} = \int \mathrm d\mu(\psi) \langle\psi|\mathcal E[\psi]|\psi\rangle\\
F_{\mathrm{ent}} = \langle \phi|(\mathbb I \otimes \mathcal E)[\phi]|\phi\rangle,
\end{cases}
\end{align}
where $F_{\min}, F_{\mathrm{avg}}$, and $F_{\mathrm{ent}}$ represent the minimum, average, and entanglement fidelity, respectively. Here, we use $\ket\phi$ to denote the maximally entangled state between the system and the ancilla (See \cref{avg fidelity pic}). Note that the entanglement fidelity ($F_{\mathrm{ent}}$) is essentially the fidelity between the Choi matrix of $\mathcal E$ and the maximally mixed state. Moreover, the measure $\mu$ in the definition of the average fidelity ($F_{\mathrm{avg}}$) is the Haar measure. In what follows, we explain how our method can be extended to compute the entanglement and the average fidelities.

Regarding the entanglement fidelity, first note that $\mathbb I \otimes \mathcal E$ is the channel that can be realized by appending an identity to the Hamiltonian and the jump operators of our Lindbladian, and then the main method of \cref{app:perturbation} can be applied with the initial state being the Choi matrix. More concretely, we get the following relations for the first-order errors
\begin{align}\label{eq:ent-gate-fid-1st-order}
\begin{split}
\epsilon_{\mathrm{ent}, L}^{(1)} &= \sum_{k} \int_{0}^{T_g} \gamma_k \bigg( \langle \mathbb I \otimes L_k^\dagger(t') L_k(t') \rangle\\
&\qquad\qquad\qquad\qquad - |\langle \mathbb I \otimes L_k(t') \rangle|^2 \bigg) \, \mathrm dt',\\
\epsilon_{\mathrm{ent}, HH}^{(2)} &= \frac{2 \delta^2}{\hbar^2}\int_{0}^{T_g} \!\!\!\int_{0}^{t'}\!\operatorname{Re}\bigg(\langle \mathbb I \otimes H_e(t) H_e(t')\rangle\\
&\qquad\qquad\quad-\langle \mathbb I\otimes H_e(t)\rangle\langle \mathbb I \otimes H_e(t')\rangle\bigg)\mathrm dt \mathrm dt',
\end{split}
\end{align}
Here, the expectations are taken with respect to the state $\ket\phi$, which is a tensor product of two $\phi^+$ Bell states, and the $\mathbb I$ operator in \eqref{eq:ent-gate-fid-1st-order}, is a $4\times 4$ operator as we consider two-dimensional registers as ancillae. Other formulas of the previous sections (such as $\varepsilon_{HL}^{(2)}$ and $\epsilon_H^{(1)}$ for the case of non-Hermitian perturbation can be generalized similarly). 

Finally, we point out that the average fidelity and the entanglement fidelity are related by the following elegant relation \cite{horodecki1999general, nielsen2002simple}
\begin{align}\label{eq:ent-to-avg}
F_{\mathrm{avg}} = \frac{DF_{\mathrm{ent}}+1}{D+1},
\end{align}
where $D$ is the effective system size (in our case $D=4$ as we are we effectively have two-qubits). Therefore, we have that the average gate infidelity $\varepsilon_{\mathrm{avg}} = 1-F_{\mathrm{avg}}$ is related to the entanglement fidelity ($\varepsilon_{\mathrm{ent}}$) via
\begin{align}
\varepsilon_{\mathrm{avg}} = \frac{D}{D+1} \varepsilon_{\mathrm{ent}}.
\end{align}

\begin{figure}
    \centering
\begin{tikzpicture}
\draw[thick, black] (-1,0) -- (1.5,0) node[midway, above] {\small ancillae};
\draw[thick, black] (-1,-.3) -- (1.5,-.3);
\draw[thick, black] (-1,-2) -- (0.1,-2);
\draw[thick, black] (-1,-2.3) -- (.1, -2.3)  node[midway, below] {system};
\draw[thick, black] (0.1, -1.5) rectangle (1.4, -2.8);
\node at (.75, -2.15) {\large gate};
\draw[thick, black] (1.4, -2) -- (2, -2);
\draw[thick, black] (1.4, -2.3) -- (2, -2.3);
\draw[thick, smooth, domain=-2:0, samples=100, color = gray] 
    plot ({-.5+0.1*sin(400*pi*\x)}, \x);
\draw[thick, smooth, domain=-2.3:-0.3, samples=100, color = gray] 
    plot ({-.8+0.1*sin(400*pi*\x)}, \x);

\draw[thick, black] (2.5,-.5) -- (3.5,-0.5);
\draw[thick, black] (2.5,-1.5) -- (3.5,-1.5);
\draw[thick, smooth, domain=-0.5:-1.5, samples=100, color = gray] 
    plot ({3+0.1*sin(400*pi*\x)}, \x);
\node at (5, -1) {\large $= \frac{1}{\sqrt2}(\ket{\uparrow\uparrow}+\ket{\downarrow\downarrow})$};
\end{tikzpicture}
\caption{The circuit computing the Choi matrix. Note that the ancillae are $2$ dimensional registers and that the squiggling lines demonstrate the Bell state $\ket{\phi^+} = \frac1{\sqrt2}(\ket{\uparrow\uparrow}+\ket{\downarrow\downarrow})$. The entanglement fidelity ($F_{\mathrm{avg}}$) of a gate is the fidelity between the Choi matrices of the ideal channel and its implementation. A simple formulation is provided in \eqref{eq:gate-fids}. Moreover, one can readily translate entanglement fidelity to average fidelity via \eqref{eq:ent-to-avg}.}
\label{avg fidelity pic}
\end{figure}

As an example, we compute the average gate fidelity of the magnetic dipolar interaction gate scheme, as shown in \cref{comparison of fids}. Additionally, we compare the state fidelity and average gate fidelity for this scheme, both obtained using the perturbative method.\\

\section{State fidelity calculation of magnetic dipolar interaction gate scheme}\label{magnetic}

\begin{figure}
\centering
\includegraphics[width=0.5\textwidth]{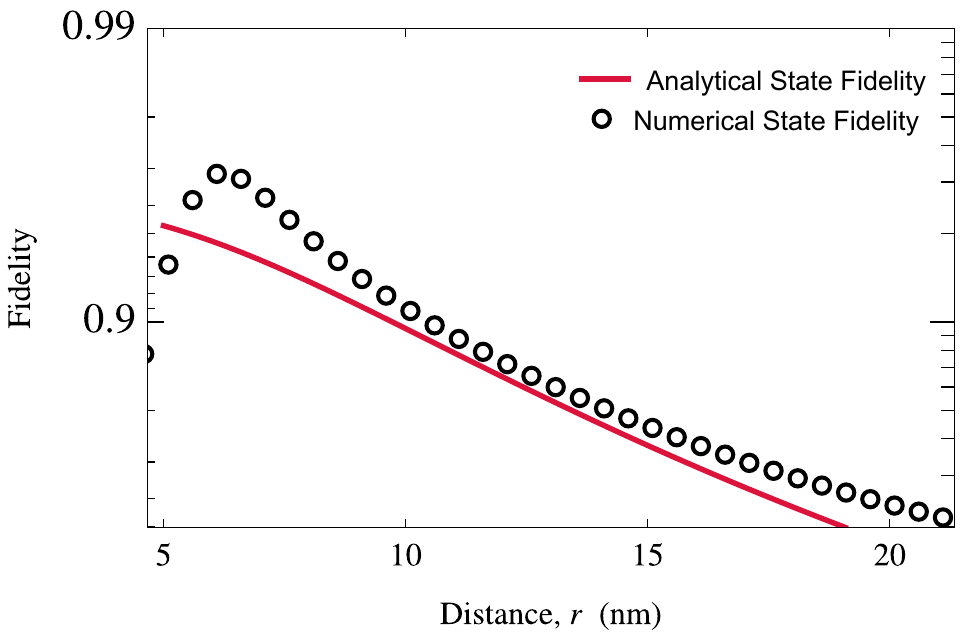}
\caption{State fidelity of the magnetic dipolar interaction gate as a function of the distance between ions. The analytic approximation using time-dependent perturbation theory including first and second-order errors (red curve) matches closely with the numerical simulation of the exact master equation dynamics (black circles). The numerical simulation, based on the parameters provided in \cref{impl:MD} of the main text, indicates that the maximum achievable fidelity, approximately 0.97, occurs at around $r \sim 6$ nm.} 
\label{MD-state-fids}
\end{figure}

To compute the fidelity of the magnetic dipolar gate, we employ the perturbative method outlined in \cref{sec:state-fid}, with the detailed derivation provided in \cref{app:perturbation}. This involves solving the Schrödinger equation to derive the ideal gate evolution and evaluating the lowest-order error expressions.
We also numerically solve the full system's master equation for a realistic set of parameters to verify the analytic approximation. \cref{MD-state-fids} compares the two solutions and demonstrates that our approach effectively captures the infidelities as expected. 

The magnetic dipole-dipole interaction phase gate can be broken down into three steps: activation, interaction, and deactivation. To obtain an analytic solution for the state evolution for the ideal gate, we assume that the activation (deactivation) step is performed using a square pulse with a temporal width $T_{\mathrm{act}}$ ($T_{\mathrm{deact}}=T_{\mathrm{act}}$) that is much shorter than the interaction time $T_\mathrm{int}$. This allows us to separate the evolution over the gate time $T_{g,MD} = 2T_\mathrm{act} + T_\mathrm{int}$ into three piecewise time-independent operations that can each be solved analytically. That is, we assume that the ions are essentially decoupled during the activation and deactivation steps.

The total Hamiltonian of the system is given by
\begin{equation}
H(t) = H_0 + H_{\mathrm{int}} + H_P(t),
\end{equation}
where
\[H_0=\sum\limits_n{\left(\omega _{\uparrow}\ket{\uparrow}\bra{\uparrow} + \omega _{\uparrow'}\ket{\uparrow'}\bra{\uparrow'}+\omega _{\downarrow'}\ket{\downarrow'}\bra{\downarrow'}\right)}_n\]
is the free Hamiltonian with $n = \{ \mathrm{Yb}_1,\mathrm{Yb}_2\}$. In the rotating frame defined by $H_0$, the driving Hamiltonian becomes time-independent $H_P^\prime = \sum\limits_n {{{\left(\frac{{{\Omega _ \Uparrow }}}{2}(\sigma _ \Uparrow + \sigma _ {\Uparrow'}) + \frac{{{\Omega _ \Downarrow }}}{2}(\sigma _ \Downarrow  + \sigma _ {\Downarrow'})\right)}_n}}$, where
\begin{align*}
\begin{split}
\sigma_{\Uparrow} &= \ket{\uparrow'}\bra{\uparrow},
\sigma_{\Uparrow'}=(\sigma_{\Uparrow})^\dag,\\
\sigma_{\Downarrow} &= \ket{\downarrow'}\bra{\downarrow},\sigma_{\Downarrow'}=(\sigma_{\Downarrow})^\dag.
\end{split}
\end{align*}
Here subscripts $\Uparrow,\Downarrow$ indicate transitions $\ket{\uparrow}\mapsto\ket{\uparrow'}$ and $\ket{\downarrow}\mapsto\ket{\downarrow'}$, while the subscripts $\Uparrow',\Downarrow'$ represent the reverse transitions $\ket{\uparrow'}\mapsto\ket{\uparrow}$ and $\ket{\downarrow'}\mapsto\ket{\downarrow}$ respectfully.
Without loss of generality, we assume
$(\Omega_{\Uparrow,\Downarrow})_n \geq 0$.
In this rotating frame, the interaction Hamiltonian remains unchanged and is governed by the magnetic dipole-dipole interaction between two Yb ions (labelled with 1 and 2), as

\begin{align}
\begin{split}
H_{\mathrm{int}} &= \frac{\mu _0}{4\pi r^3} \left[ {\mathbf{\mu}}_1\cdot{\mathbf{\mu}}_2 - \frac{3\left({ \mathbf{\mu} }_1\cdot\mathbf{r} \right)\left({\mathbf{\mu} }_2\cdot\mathbf{r} \right)}{r^2} \right],
\end{split}
\end{align}
where $\mu_0$ is the vacuum permeability, $\mathbf{\mu}$ is the electronic magnetic dipole moment which is defined as $\mathbf{\mu}=\mathbf{\mu_B\cdot g\cdot S}$, with $\mu_B$ being the Bohr magneton, and $\mathbf{g}$ is g tensor with principal values of $g_i$ (for $i = x,y,z$). Here $\mathbf{r}$ is the distance between two ions in the crystal, where we assume two ions are located along the z-axis ($c$ axis of the YVO crystal). In cases of imperfect alignment along the z-axis, additional error terms arise. However, these coherent errors stem from a Hamiltonian associated with a unitary evolution. As a result, these errors can, in principle, be mitigated by applying local operations to each qubit. We consider the point symmetry with $g_x\simeq g_y \equiv g_{\perp}$, and $g_z\equiv g_{\|}$. While $g_{\perp}$ governs the transverse parts of the interaction (such as the spin flip-flop process), we compute the errors arising from these transverse interactions in detail. Therefore, we divide the interaction Hamiltonian into two parts
\begin{equation}\label{interaction}
H_{\mathrm{int}}= H_{I}+H_{e},
\end{equation}
with
\begin{align*}
\begin{split}
{H_I} = J_{\|}\,  Z_1 \otimes Z_2,\quad
H_e = J_x\, X_1\otimes X_2 + J_y\, Y_1\otimes Y_2, 
\end{split}
\end{align*}
where $H_I$ is an Ising-type interaction with coupling strength $J_{\|}=\frac{-\mu _0(\mu _Bg_{\|})^2}{8\pi r^3}$ and $X,Y,Z$ represent Pauli matrices. The transverse interactions are given with $H_e$, where
$J_i=\frac{\mu_0(\mu_Bg_i)^2}{16\pi r^3}$ (for $i=x,y$) is the transverse interaction strength.\\

The reduced magnetic dipole-dipole interaction (denoted by $H_I$) can be exploited to execute the gate by use of the non-trivial unitary
evolution of the active qubits $(U_\mathrm{I}(t)=e^{-i H_I t})$. The activation and deactivation steps can be approximated by the evolution $U_P(t)=e^{-i H_P^\prime t}$ in the rotating frame. With these pieces, we can write the total gate propagator (i.e., the unitary) describing the evolution as follows
\begin{equation}
\label{eq:t_act-unitary}
U_g(t)=
\left\{
\begin{aligned}
&U_P(t)\\
&\qquad \qquad 0 \le t \le T_{\mathrm{act}},\\[2pt]
&U_I(t-T_{\mathrm{act}})\,U_P(T_{\mathrm{act}})\\
&\qquad \qquad 0 < t-T_{\mathrm{act}} \le T_{\mathrm{int}},\\[2pt]
&U_P(t-T_{\mathrm{act}}-T_{\mathrm{int}})
   \,U_I(T_{\mathrm{int}})\,U_P(T_{\mathrm{act}})\\
&\qquad \qquad t > T_{\mathrm{act}}+T_{\mathrm{int}}.
\end{aligned}
\right.
\end{equation}
Therefore, the final state after the evolution is
\begin{equation}
\ket{\psi(T_g)} = U_g(T_g)\ket{\psi(0)}.
\end{equation}

For this gate analysis, the sources of error are the transverse dipole-dipole interaction described by the Hermitian Hamiltonian $H_e$ and any decoherence processes such as spontaneous emission and dephasing. Thus, we can immediately conclude that $\epsilon_H^{(1)}=0$ and focus on the first-order term $\epsilon_L^{(1)}$ and the second-order term $\epsilon^{(2)}_{HH}$.

To evaluate the irreversible error $\epsilon^{(1)}_L$, we assume identical optical spontaneous-emission rates for the two ions, denoted by $\gamma_{1,\Uparrow'}$ ($\gamma_{1,\Downarrow'}$), as well as identical bulk optical pure-dephasing rates $\gamma_{2,\Uparrow'}$ ($\gamma_{2,\Downarrow'}$). These processes induce decay and dephasing on the optical transitions $\ket{\uparrow'} \rightarrow \ket{\uparrow} (\ket{\downarrow'} \rightarrow \ket{\downarrow})$. The corresponding Lindblad collapse operators are $\sigma_{\Uparrow'}$ ($\sigma_{\Downarrow'}$) for spontaneous decay, and $\mathrm{diag}\{-\tfrac{1}{2}, -\tfrac{1}{2}, \tfrac{1}{2}, \tfrac{1}{2}\}$ for pure dephasing.
We also consider a spin relaxation rate $\gamma_{3}$ ($\gamma_{4}$) that refers to the ground (excited) state spin-flip corresponding to the Lindblad operator $\ket{\uparrow}\bra{\downarrow}$ ($\ket{\uparrow^\prime}\bra{\downarrow^\prime}$) and a spin pure dephasing rate $\gamma_{5}$ ($\gamma_{6}$) of the ground (excited) state spin doublet corresponding to Lindblad operator $\ket{\downarrow}\bra{\downarrow} - \ket{\uparrow}\bra{\uparrow}$ ($\ket{\downarrow^\prime}\bra{\downarrow^\prime} - \ket{\uparrow^\prime}\bra{\uparrow^\prime}$).
Putting these together, we evaluate $\epsilon^{(1)}_L$ given in \cref{eps-first-order} using the ideal gate propagator in \cref{eq:t_act-unitary} to arrive at
\begin{align}\label{e1-MD}
\begin{split}
\varepsilon _L^{(1)}=&
{T_\mathrm{act}}\bigg(\frac{{7}}{{8}}({\gamma _{1, \Uparrow '}} + {\gamma _{1, \Downarrow '}}) + 
\frac{1}{2}(\gamma_{2,\Uparrow'}+\gamma_{2,\Downarrow'})\\
&\qquad+\frac{{13}}{{16}}({\gamma _2} + {\gamma _3}) + \frac{1}{{2}}({\gamma _4} + {\gamma _5})\bigg)  \\ 
 +&{T_{{\mathop{\rm int}} }}\left({\gamma _{1, \Uparrow '}} + {\gamma _{1, \Downarrow '}} + \frac{{3}}{{4}}{\gamma _3} + \frac{{1}}{{2}}{\gamma _5}\right),
 \end{split}
\end{align}
where we have assumed $\Omega_{\Uparrow}=\Omega_{\Downarrow} = \Omega=\pi/T_\mathrm{act}$ and $T_\mathrm{int}=\hbar\pi/4J_{\|}$.

For the reversible error, we evaluate \cref{eps-second-order} using the error Hamiltonian given by $H_e$ in \cref{interaction} along with the ideal unitary propagator to obtain

\begin{equation}\label{HH_MD}
\epsilon _{HH}^{(2)} =\frac{a(J_{x}+J_{y})^2}{{J_{z}^2}},
\end{equation}
where the coefficient is $a = (32(2-\sqrt{2}) - {\pi ^2})/64$. For clarity, we add that the ideal Hamiltonian evolution considered in the computation of \cref{HH_MD}, is the $Z\otimes Z$ interaction term together with the free Hamiltonian. In the absence of free Hamiltonian, the expression would be much simpler as the error term $X\otimes X + Y\otimes Y$ commutes with $Z\otimes Z$. However, due to the presence of the free Hamiltonian, we end up with the more complex pre-factor $a$ in \cref{HH_MD}.
Substituting the Yb system's parameters, we find that the transverse interaction strength, $J_x = J_y \equiv J_{\perp}$, is 4.35 times weaker than the longitudinal interaction strength, $J_z \equiv J_{||}$. Consequently, the contribution of second-order errors in \cref{HH_MD} to the state fidelity equation is approximately $10^{-2}$, supporting the validity of our assumption that it can be treated perturbatively.

Furthermore, \cref{MD-state-fids} presents a comparison between the fidelities obtained by simulating the exact system (i.e., solving the master equation) and those obtained from our perturbative solution. This comparison confirms that the two approaches yield consistent results within the expected ranges of $r$.
\cref{MD-state-fids} provides additional insights into the ion separation regime of interest for the Yb:YVO system. Firstly, at small $r$, the dipolar interaction affects the excitation pulse in a non-trivial way which is not captured by our analytical approach. This complex behavior, however, is evident in the numerical simulation.
Note that the dipolar gate's non-trivial effect can be safely ignored when $\norm{H_{\mathrm{int}} T_{\mathrm{act}}} \ll 1$, which corresponds to large enough $r$ (e.g., $r\gtrsim  5 \mathrm{nm}$ with the existing parameters for Yb:YVO system). Secondly, for sufficiently large distances (e.g., greater than 20 nm for the system under study), the two solutions diverge. This occurs because the perturbative approach does not hold in the low-fidelity regime corresponding to longer interaction times (we recall that according to the perturbation solution in \cref{f_md} the fidelity scales as $\propto 1-O(r^3)$ with interaction time $T_{\mathrm{int}}\propto r^3$).

\section{Comparison between state fidelity and average gate fidelity of magnetic dipolar gate scheme utilizing perturbative method}\label{comparison of fids}

\begin{figure}
\centering
\includegraphics[width=0.5\textwidth]{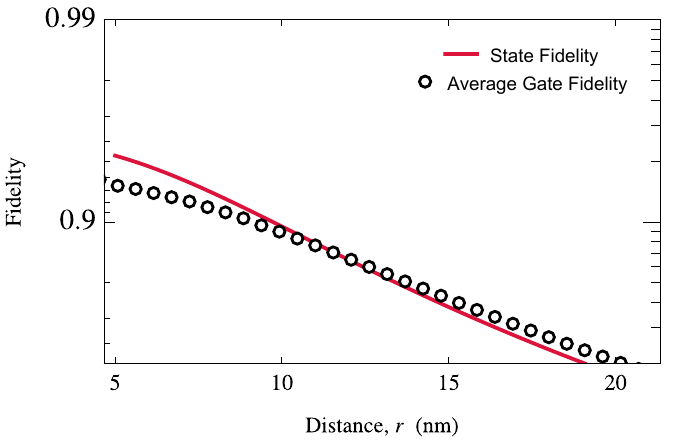}
\caption{Analytical state fidelity (red curve) and the average gate fidelity (black curve) of the magnetic dipolar interaction two-qubit phase gate as a function of ions' separation. We note that although the average gate fidelity is also obtained with the time-dependent perturbative method, we solve the integrals numerically, as the system size for this simulation is much larger (we have a $64$-dimensional space), making symbolic computation extremely difficult. The parameters utilized are detailed in \cref{impl:MD} of the main text.} 
\label{fig:gate-fid}
\end{figure}

Here we compute the average gate fidelity of the magnetic dipolar interaction scheme (introduced in \cref{sec:magnetic-dipolar}), utilizing the equations presented in \cref{gate-fids}. The results of this calculation are brought in \cref{fig:gate-fid}, where we compare the state fidelity against the average gate fidelity. We observe that the two fidelities are comparable and have similar behaviour in the considered range of parameters. \\

\section{Implementation of a controlled-Z gate based on photon interference-based scheme}\label{barret-kok appendix}

Following \cite{lim2005repeat}, one can apply a CZ gate using different measurement setups. Let the first and the second ion be initially described by the arbitrary joint state
\begin{align}
\begin{split}
\ket{\psi_{\mathrm{init}}} =& \alpha \ket{\uparrow_{\mathrm{Yb}_1} \uparrow_{\mathrm{Yb}_2}} + \beta\ket{\uparrow_{\mathrm{Yb}_1}\downarrow_{\mathrm{Yb}_2}}\\
+& \gamma \ket{\downarrow_{\mathrm{Yb}_1}\uparrow_{\mathrm{Yb}_2}} + \delta \ket{\downarrow_{\mathrm{Yb}_1}\downarrow_{\mathrm{Yb}_2}}.
\end{split}
\end{align}
After excitation $\ket{\uparrow}\mapsto\ket{\uparrow'}$ and spontaneous emission steps, we get the following joint entangled state of the ions and photons
\begin{align}
\begin{split}
\ket{\psi_{\mathrm{ent}}}=& \alpha \ket{\uparrow_{\mathrm{Yb}_1} \uparrow_{\mathrm{Yb}_2}}\ket{1_{\mathrm{Yb}_1} 1_{\mathrm{Yb}_2}}\\ &+ \beta\ket{\uparrow_{\mathrm{Yb}_1}\downarrow_{\mathrm{Yb}_2}}\ket{1_{\mathrm{Yb}_1} 0_{\mathrm{Yb}_2}}\\
&+ \gamma \ket{\downarrow_{\mathrm{Yb}_1}\uparrow_{\mathrm{Yb}_2}}\ket{0_{\mathrm{Yb}_1} 1_{\mathrm{Yb}_2}}\\ &+ \delta \ket{\downarrow_{\mathrm{Yb}_1}\downarrow_{\mathrm{Yb}_2}}\ket{0_{\mathrm{Yb}_1} 0_{\mathrm{Yb}_2}}.
\end{split}
\end{align}
Instead of measuring in the number basis of photons, we can measure in a mutually unbiased basis (MUB), where the measurement projects onto states of the following form of the cavities
\begin{align}
\begin{split}
\ket{\Phi_{\mathrm{meas}}} &= \frac12\bigg(e^{i\phi_1} \ket{0_{\mathrm{Yb}_1} 0_{\mathrm{Yb}_2}} + e^{i\phi_2} \ket{0_{\mathrm{Yb}_1} 1_{\mathrm{Yb}_2}}\\
&+ e^{i\phi_3} \ket{1_{\mathrm{Yb}_1} 0_{\mathrm{Yb}_2}} + e^{i\phi_4} \ket{1_{\mathrm{Yb}_1} 1_{\mathrm{Yb}_2}}\bigg).
\end{split}
\end{align}
Note that
\begin{align}
\begin{split}
(\mathbb I \otimes \bra{\Phi_{\mathrm{meas}}}) \ket{\psi_{\mathrm{ent}}} &\propto e^{-i\phi_1} \alpha \ket{\uparrow_{\mathrm{Yb}_1} \uparrow_{\mathrm{Yb}_2}}\\ &+ e^{-i\phi_2} \beta \ket{\uparrow_{\mathrm{Yb}_1} \downarrow_{\mathrm{Yb}_2}}\\
&+ e^{-i\phi_3} \gamma \ket{\downarrow_{\mathrm{Yb}_1} \uparrow_{\mathrm{Yb}_2}}\\ &+ e^{-i\phi_4} \delta \ket{\downarrow_{\mathrm{Yb}_1} \downarrow_{\mathrm{Yb}_2}}.
\end{split}
\end{align}
Hence, post selecting on the state of the cavities being $\ket{\Phi_{\mathrm{meas}}}$, we are effectively applying the gate $\mathrm{diag}(e^{-i\phi_1},e^{-i\phi_2},e^{-i\phi_3},e^{-i\phi_4})$. Choosing $\phi_1=\phi_2=\phi_3=0$ with $\phi_4=\pi$ implements the controlled-Z gate.

\section{Effect of dark and noise counts in the photon interference-based scheme}\label{dark}

To specifically address the dark count effect in the photon interference–based scheme, we provide the following analysis. 
By expanding \cref{f_bk_cavity} of the main text, which represents the fidelity of this scheme, with respect to $1/F_p$ around $1/F_p=0$, we obtain

\begin{equation}\label{fp_large}
    F'_\text{PI} = 1 - \frac{2\gamma^{\ast}}{F_p\gamma_r},
\end{equation}
where $\gamma^{\ast}$ denotes the optical pure dephasing rate of Yb ions in cavity, $F_p$ is the Purcell factor, and $\gamma_r$ is the radiative decay rate. We also use $\varepsilon_{\mathrm{PI}}=1-F'_{\mathrm{PI}}$ to denote the infidelity associated with this error.

The probability of dark count events can be expressed as
\begin{equation}
    P_\text{d} = 7 T'_{1p} R,
\end{equation}
where $R$ is the dark count rate. Also $7T'_{1p}$ is the
detection window, where $T'_{1p}$ denotes the Purcell-enhanced
lifetime of the photon set by the emitter, which is fixed by $T_{1o,\mathrm{cav}}$.
The Purcell enhanced lifetime is defined to be $T_{1o,\mathrm{cav}}=\frac{1}{\gamma'}=\frac{1}{\gamma_{1r}F_p+\gamma_1}$. In the limit of large $F_p$ we have $T_{1o,\mathrm{cav}}\approx\frac{1}{\gamma_rF_p}$.
The factor 7 arises because, in photon detection-based schemes, the photon truncation time sets the total gate duration.  
Stopping detection too early reduces efficiency, while waiting too long increases spin decoherence.  
Our simulation indicates that choosing a truncation time of $7T'_{1p}$ minimizes fidelity loss to about $0.1\%$ across the explored cooperativity range in the main text.

In the photon interference-based scheme, the probability of detecting two heralding photons is $\eta^2/2$, where $\eta$ is the total single-photon detection probability. In the regime where the probability of detecting a signal photon is much larger than that of a dark count (i.e., $\eta \gg P_d$), the dominant false coincidence count arises from events in which one signal photon is detected, and the other detection event is caused by a dark count.

Therefore, we define the noise-to-signal ratio (NSR) to be

\begin{equation}
    \mathrm{NSR}
    =
    \frac{2\eta(1-\eta)\times 4 P_d (1-P_d)^3 / 2}
         {\eta^2 (1-P_d)^4 / 2}.
\end{equation}\label{SNR_def}
Here, $2\eta(1-\eta)$ is the probability that two photons are emitted but only one is successfully detected; the factor of $2$ accounts for the two possible choices of the lost photon. The factor $4 P_d (1-P_d)^3$ is the probability that exactly one of the four detection windows (corresponding to two detectors and two rounds of emission) produces a dark count, while the other three do not. Assuming a balanced setup, the factors of $1/2$ in both numerator and denominator arise from post-selection.

In the high-loss regime ($\eta \ll 1$), together with low dark-count probability and under the assumption $\eta \gg P_d$, this simplifies to
\begin{equation}\label{SNR}
    \mathrm{NSR}
    \approx
    \frac{8 \eta P_d}{\eta^2}
    =
    \frac{8 P_d}{\eta}
    =
    \frac{8 \times 7 R}{\gamma_rF_p \eta}.
\end{equation}

In this case, the fidelity can then be bounded as
\begin{equation}
    F \geq F'_{\mathrm{PI}} (1-\mathrm{NSR})
    =
    (1-\varepsilon_{\mathrm{PI}})(1-\mathrm{NSR})
    \approx
    1-\varepsilon_{\mathrm{PI}}-\mathrm{NSR},
\end{equation}
where $F'_{PI}$ is the fidelity of the scheme prior to considering the dark count effect, and in the limit of large $F_p$ is given by \cref{fp_large}.
To ensure that the dark-count contribution does not dominate, we should have
\begin{equation}
    \varepsilon_{\mathrm{PI}} > \mathrm{NSR}.
\end{equation}
Using our estimates in \cref{fp_large} and \cref{SNR} we have
\begin{equation}
    \frac{2\gamma^*}{F_p \gamma_r}
    >
    \frac{8 \times 7  R}{\gamma_rF_p \eta},
\end{equation}
which implies
\begin{equation}
    \frac{28R}{\eta \gamma^*} < 1.
\end{equation}

Here, the coefficient $28$ arises from the specific photon-interference–based protocol and may vary for schemes. 

\begin{figure}[h]
    \centering
    \includegraphics[width=0.9\linewidth]{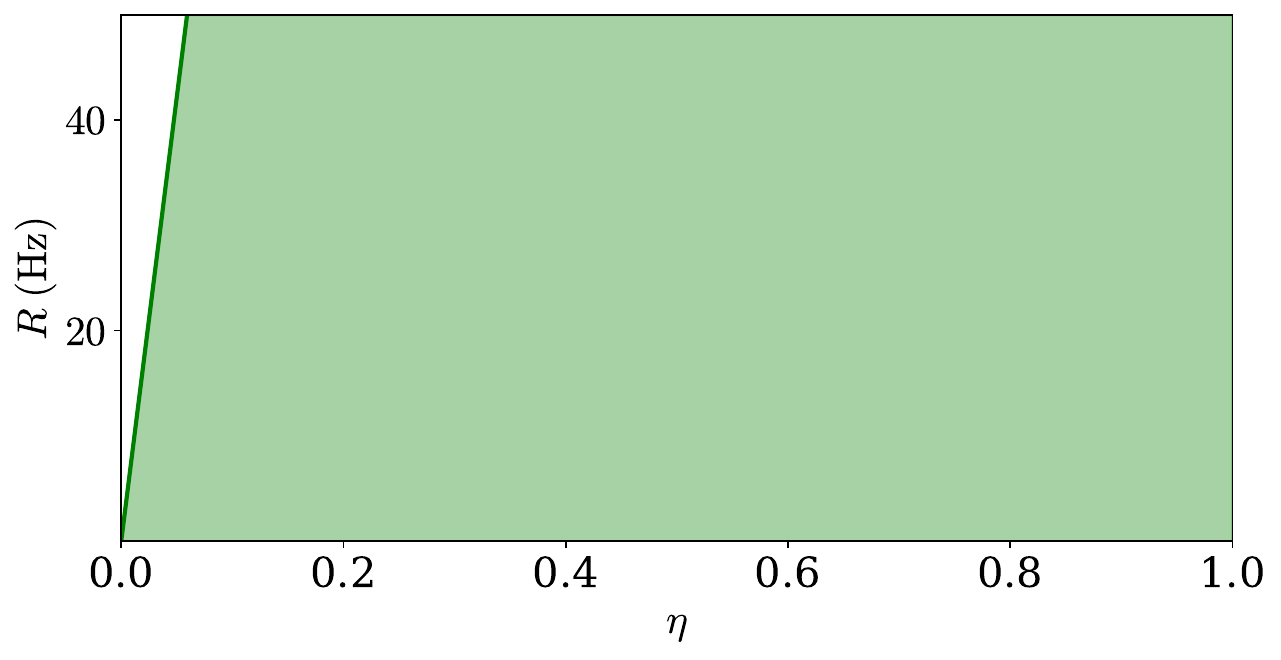}
    \caption{Ratio of the dark count error to the gate dephasing error. The green region indicates the parameter space where this ratio remains below unity, meaning that dark counts are not the limiting factor.}
    \label{dark counts}
\end{figure}

Importantly, since the probabilities of dark count errors and gate errors are both inversely proportional to the Purcell factor, the gate fidelity can, in principle, be arbitrarily improved by increasing the Purcell factor.
For example, based on current experimental parameters ($\eta \sim 1.1\times10^{-2}$~\cite{ruskuc2025multiplexed}, optical pure dephasing in the cavity $\gamma^{\ast} \approx 23.64~\text{kHz}$), and considering $R \approx 9~\text{Hz}$ this ratio is approximately unity. 
Increasing the cavity cooperativity can help suppress both gate error and dark count contributions. 
\cref{dark counts} is also presented to illustrate the parameter regimes in which this ratio exceeds unity, showing where dark counts become the dominant error source. 
This plot indicates that for moderately good efficiency, dark counts are not the dominant limitation, highlighting the critical role of efficiency enhancements.

\section{Fidelity calculation of photon scattering gate scheme}\label{app: photon scattering}

Our method for computing the fidelity of the photon scattering scheme is based on the so-called SLH formalism \cite{combes2017slh}, which stands for Scattering matrix, Lindblad operators, and Hamiltonian. This formalism allows to easily construct quantum master equations that include cascaded interactions mediated by waveguide modes. In order to obtain the master equation, we consider a cascaded system where a single photon source is modelled by an ideal two-level emitter whose emission is cascaded into the cavity that contains both ions. In the single-excitation regime, the cavity can additionally be described by a two-level system. For the ions, since we assume one of the two transitions is far off resonant with the cavity, we approximate each of them using a three-level system. Thus, the total Hilbert space of the simulated system has a size $2\times2\times3\times3 = 36$, corresponding to a Fock-Liouville space of $36^2=1296$.

The evolution of the incoming single-photon pulse can be described by an SLH triple $(S_s, \mathbf L_s, H_s)$, where $S_s = \mathbb I$ is the identity operator, and $H_s = 0$ is the cavity Hamiltonian (here $s$ stands for `source').  The Lindblad operators $\mathbf L_s$ include $\sqrt\Gamma_s a_s$ as the only non-zero component, where $\Gamma_s=1/T_{1p}$ is the scattered photon bandwidth and $a_s$ is the source photon annihilation operator. The cavity-ion system is modelled by the SLH triple $(S_0, \mathbf L_0, H_0)$ where $S_0=\mathbb I$ and $H_0= \sum\limits_n{\omega _n} \sigma _{ \uparrow n}^\dag {\sigma _{ \uparrow n}} + ({\omega _n} + {\omega _e})\sigma _{ \downarrow n}^\dag {\sigma _{ \downarrow n}} + {\omega _g}\sigma _{ \uparrow  \downarrow n}^\dag {\sigma _{ \uparrow  \downarrow n}}$, where $\omega_n$ is the frequency between $\ket{\uparrow}$ and $\ket{\uparrow'}$, $\omega_e$ is the separation frequency between $\ket{\uparrow'}$ and $\ket{\downarrow'}$ in the excited state, and $\omega_g$ is the separation frequency between $\ket{\uparrow}$ and $\ket{\downarrow}$ in the ground state.
The Lindblad operators are $\sigma_{\uparrow}\ket{\uparrow'}=\ket{\uparrow}$, $\sigma_{\downarrow}\ket{\downarrow'}=\ket{\downarrow}$, and $\sigma_{\uparrow\downarrow}\ket{\downarrow}=\ket{\uparrow}$.
The collapse operators $\mathbf L_0$ corresponding to the emission of light are $(\sqrt{\kappa}a_c, \sqrt{\gamma_{1,\Uparrow}}\sigma_{\uparrow}, \sqrt{\gamma_{1,\Downarrow}}\sigma_{\downarrow})$
and we additionally include optical and spin decoherence modeled by the collapse operators $\sqrt \gamma^\star \sigma_{\uparrow}^{\dag}\sigma_{\uparrow}$, $\sqrt \gamma_2 \sigma_{\uparrow,n}$, and $\sqrt \gamma_4 \sigma_{\uparrow,n}^\dagger \sigma_{\uparrow,n}$ for both ions. 

Using the cascade rules from \cite{combes2017slh}, we have that the final triple is $(\mathbb I, \mathbf L, H)$, where $\mathbf L = \mathbf L_0 + \mathbf L_s$, and $H = H_0 + H_s + V$ with $V$ being the cascaded interaction term. This final triple then defines a time-independent master equation $\dot{\rho} = \mathscr{L}\rho$, which is then solved by $\rho(t) = e^{t\mathscr{L}}\rho(0)$ for an initial state $\rho(0)=\rho_s(0)\otimes\rho_c(0)\otimes \rho_1(0)\otimes\rho_2(0)$ where $\rho_s(0)=\ket{1}\bra{1}$, $\rho_c(0)=\ket{0}\bra{0}$, $\rho_i(0)=\ket{\psi(0)}\bra{\psi(0)}$, and $\ket{\psi(0)}=(\ket{\uparrow} + \ket{\downarrow})/\sqrt{2}$.

Since we obtain the time-dependent solution (the virtual source cavity is not driven), the resulting photon that cascades into the cavity containing the ions will have a Lorentzian spectral shape. This method could be used along with a time-dependent $\Gamma_s(t)$ to shape the photon, but at the cost of a significantly increased simulation time.

On the other hand, a Lorentzian-shaped photon complicates the analytic solution using the method described in \cite{asadi2020cavity}. This is because Lorentzian shapes diverge under integration, and thus require a truncation. In any case we expect that in \cref{fid:PS}, at least for $\alpha\ll 1$, only the relationship between $\sigma_p$ and $T_g$ depends on the exact shape of the photon.

In principle, it should be possible to obtain an analytic approximation of the fidelity directly from the cascaded master equation, perhaps using a modification of the perturbative approach but for cascaded systems. However, we have thus far failed to do so because the ideal gate evolution is irreversible. This could be a topic for future exploration.

\section{Fidelity calculation of virtual photon exchange gate scheme}\label{fid:virtual} 

The fidelity of the virtual photon exchange scheme was previously established by some of us in Ref \cite{asadi2020cavity}. 
However, the method used could not account for all dissipative parameters, which resulted in an upper bound on gate fidelity.
Later, the authors in Refs \cite{asadi2020protocols, wein2021modelling} used numerical computation to add the effect of optical pure dephasing rate to the fidelity. Using the theoretical framework outlined in \cref{sec:state-fid} and assuming that the system operates in the bad cavity limit and considering the adiabatic regime, we reproduce these results and provide an analytical formulation for the fidelity of the virtual photon exchange scheme.

In the case of virtual photon exchange, it is useful to model noise using an effective non-Hermitian Hamiltonian. This is because the ideal gate evolution arises due to the adiabatic elimination of the mediating cavity mode. From this perspective, the cavity decay rate causes a non-Hermitian component to arise in the effective ion-ion interaction Hamiltonian.

We begin by writing the total Hamiltonian for the three-body system
\begin{equation}
\label{3bodyH}
   H =  H_0  +  H_C+ H_I,
\end{equation}
where
\begin{align}
\begin{split}
H_0 &= \sum\limits_n{\omega _n} \sigma _{ \uparrow n}^{\dag} {\sigma _{ \uparrow n}} + ({\omega _n} + {\omega _e})\sigma _{ \downarrow n}^{\dag} {\sigma _{ \downarrow n}} + {\omega _g}\sigma _{ \uparrow  \downarrow n}^\dag {\sigma _{ \uparrow  \downarrow n}},\\
H_C &= {\omega _C}{a^\dag }a,\\
H_I &= \sum_n {\sum_k {{g_{nk}}} } (\sigma _{nk}^\dag a + h.c.).
\end{split}
\end{align}
Here $g_{\uparrow,n}$ is the cavity coupling rate of transition $(\ket{\uparrow}\mapsto\ket{\uparrow'})_n$, $g_{\downarrow,n}$ is the cavity coupling rate of the transition $(\ket{\downarrow}\mapsto\ket{\downarrow'})_n$. We also assume $g_{nk}=g$ is the same for both transitions in both Yb ions.

Let $\Delta_n=\omega_n-\omega_c$ be the detuning between the cavity and the $n^{\text{th}}$ ion's optical transition $\ket{\uparrow}\mapsto\ket{\uparrow'}$. In the high-cooperativity regime, a large cavity detuning $\Delta=\Delta_{\mathrm{Yb}_1}\simeq\Delta_{\mathrm{Yb}_2}+\delta_{eg}$ allows for a dispersive interaction to mediate the two-qubit gate. In this regime, the unitary operator describing the evolution can be solved by adiabatically eliminating the cavity mode to obtain the ion-ion interaction Hamiltonian $H_\mathrm{AE}$ valid when $\delta_{eg}\gg\Delta\gg g$ \cite{asadi2020cavity}. In the limit that $\Delta\rightarrow \infty$ and $\delta_{eg}\rightarrow\infty$, a control phase gate is perfectly implemented after an interaction time of $T_g=\pi\Delta/g^2$.

The true time evolution operator in the absence of irreversible errors is given by $U(t)=e^{-i\hat H t}$ and the final state is $\ket{\psi(t)}=U(t)\ket{\psi(0)}$. However, to use the perturbative method, it is necessary to analytically solve the system governed by the perfect Hamiltonian $H_g=\lim_{\Delta,\delta_{eg}\rightarrow\infty}H_\mathrm{AE}$ with an error Hamiltonian $H_e=H-H_g$. Due to the large detunings, the eigenvalues of $H_e$ can be much larger than those of $H$, which violates the conditions to apply perturbation theory. Thus, it is necessary also to eliminate the cavity mode for the error Hamiltonian and use $H_e=H_{AE}-H_g$. Unfortunately, eliminating the cavity mode in this way also eliminates the direct cavity-emitter coupling and hence all impact due to the cavity decay is lost, which greatly modifies the qualitative features of the gate error.

To maintain a dependence on the cavity decay rate while deriving a valid perturbation, we first consider the effective non-Hermitian Hamiltonian $H_\mathrm{eff}=H-i\frac{\kappa}{2}a^{\dag}a$ of the three-body system. This effective Hamiltonian captures the amplitude damping effect due to the cavity in the Hamiltonian evolution, but neglects the stochastic jump of the cavity mode to the ground state. Luckily, such a jump forces the system to leave the expected computational space and thus can be neglected without affecting the state fidelity \cite{asadi2020cavity}.

We then proceed to eliminate the cavity mode for the non-Hermitian Hamiltonian $H_\mathrm{eff}$ to obtain the error-prone interaction Hamiltonian $H_\mathrm{eff, AE}$. The damping effect caused by the cavity decay rate on the effective coupling rate now appears as a non-Hermitian error Hamiltonian $\tilde H_e=H_{g}-H_\mathrm{eff, AE}$. With this, we can evaluate the expressions in \cref{eps-first-order} to obtain the non-zero first-order error terms while accounting for any additional decoherence terms impacting the emitter systems. Notably, unlike the method used in Ref. \cite{asadi2020cavity}, this approach allows for the analytic evaluation of first-order errors due to optical pure dephasing of the emitters.

Following this procedure, we find
\begin{equation}
\begin{aligned}
    \epsilon_L^{(1)}&=\frac{\pi\kappa}{2\Delta}+T_g\gamma_6+\frac{21}{32}T_g\gamma^{*}\\
    \epsilon_H^{(1)}&=\frac{2\pi\Delta}{C\kappa},
\end{aligned}
\end{equation}
where $\gamma^{*}$ is optical dephasing rate equal for both ions, $\Delta$ is cavity detuning, $C=4g^2/\kappa\gamma_1$ is cavity cooperativity, $\kappa$ is cavity decay rate, and $\gamma_6$ is spin dephasing rate of the excited state for both ions. We use this notation to maintain consistency with the cooperativity definition in the main text and the notation in \cref{e1-MD}. This result is exactly the same as the expression obtained in \cite{asadi2020cavity} (except a factor of two which comes from fidelity definition) once considering the optimal detuning condition $2\Delta=\kappa\sqrt{C}$, but now we include pure dephasing terms. Thus, the perturbative approach demonstrated in this work offers a real practical advantage, particularly in terms of faster symbolic computation times, especially in higher-dimensional Hilbert spaces. Note that the study in \cite{asadi2020protocols} also considered the impact of optical pure dephasing for this gate, but the technique relied on numerical simulations of the master equation to infer the scaling constant. Here, the perturbative approach allows one to capture the effects of all dissipative parameters up to any arbitrary order analytically.

\begin{figure}
\centering
\includegraphics[width=0.5\textwidth]{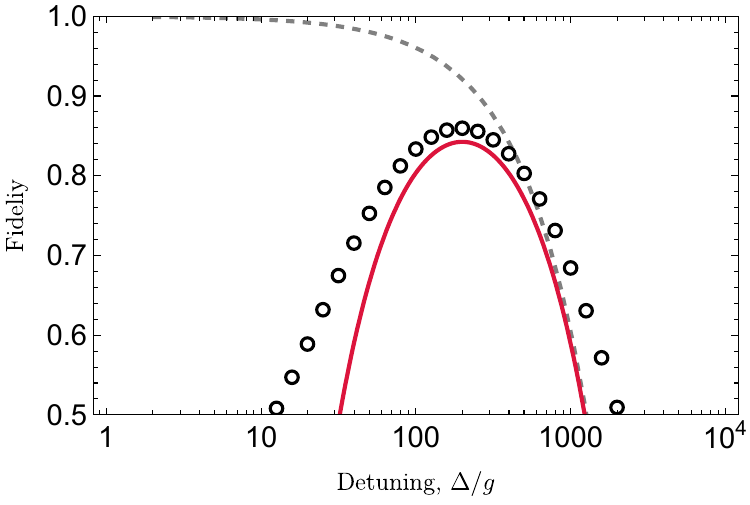}
\caption{State fidelity of the simple virtual photon exchange phase gate as a function of the cavity detuning $\Delta/g$. The analytic approximation using first-order time-dependent perturbation theory (red curve) matches closely with the numerical simulation of the exact master equation dynamics (black circles) when fidelity is more than 0.8. The gray dashed line indicates the approximated error when neglecting the impact of cavity decay. The parameters used are $\kappa/g=10$ indicating a weak cavity-emitter coupling regime, $\gamma_1 = 10^{-4}g$, $\delta_{eg}=100g$, $\gamma^{*}=10^{-4}g$, and $\gamma_6=10^{-5}g$ corresponding to a large cooperativity of $C=4\times 10^4$.} 
\label{fidelity-cavity}
\end{figure}

To validate the analytic approximation, we compare it to an exact numerical simulation of the master equation dynamics for parameters where all relevant sources of error are non-negligible (see \cref{fidelity-cavity}).

\section{Yb:YVO system's parameters}\label{app:system-parameters}
\Cref{tab:YbYVO_parameters} summarizes the parameters used in the simulations of all gate schemes investigated in this work. The bulk parameters employed for the magnetic dipolar gate are listed separately from the cavity parameters used for the photon-scattering and photon-interference-based schemes.

\begin{table*}[t]
\centering
\caption{Yb$^{3+}$:YVO$_4$ system parameters used in the simulation of the gate schemes studied in this work.}
\label{tab:YbYVO_parameters}
\renewcommand{\arraystretch}{1.2}
\footnotesize
\setlength{\tabcolsep}{5pt}
\begin{tabular}{@{}l c c@{}}
\toprule
\textbf{Parameter} & \textbf{Magnetic dipolar scheme} & \textbf{Cavity-based schemes} \\
\midrule

Optical coherence time
  & $T_{2o}=91\,\mu$s\,(at $B\!\approx\!500$\,mT)\,\cite{kindem2018characterization}
  & $T_{2o, \mathrm{cav}}=4.1\,\mu$s\,\cite{kindem2020control} \\
  Optical lifetime
  & $T_{1o}=267\,\mu$s\,\cite{kindem2018characterization}
  & $T_{1o, \mathrm{cav}}=2.27\,\mu$s\,\cite{kindem2020control} \\
Optical pure dephasing rate
  & $\gamma_2 = 1/T_{2o}\!-\!1/2T_{1o}\!\approx\!9.12$\,kHz
  & $\gamma^\ast = 1/T_{2o,\text{c}}\!-\!1/2T_{1o,\text{c}}\!\approx\!23.63$\,kHz \\
Ground state spin coherence time
  & $T_{2s,g}=6.6$\,ms\,(at $B\!=\!440$\,mT)\,\cite{kindem2018characterization}
  & $T_{2s,g,\mathrm{cav}}=31$\,ms\,\cite{kindem2020control} \\
Ground state spin lifetime
  & $T_{1s, g} =200$\,ms\,\cite{macfarlane1987coherent, kindem2018characterization}
  & $T_{1s,g,\mathrm{cav}}=54$\,ms\,\cite{kindem2020control} \\
Ground state spin dephasing rate
  & $\gamma_5=1/T_{2s,g}\!-\!1/2T_{1s,g}\!\approx\!149$\,Hz
  & Not included in simulations. \\
Excited state spin coherence time
  & $T_{2s,e}=35\,\mu$s\,(at $B\!\to\!0$)\,\cite{bartholomew2020chip}
  & " \\
Excited state spin lifetime
  & Assumed $T_{1s,e}=T_{1s,g}$
  & " \\
Excited state spin dephasing rate
  & $\gamma_6=1/T_{2s,e}\!-\!1/2T_{1s,e}\!\approx\!28.57$\,kHz
  & " \\
Cavity coupling
  & ---
  & $g=2\pi\!\times\!23$\,MHz\,\cite{kindem2020control} \\
Cavity linewidth
  & ---
  & $\kappa=2\pi\!\times\!30.7$\,GHz\,\cite{kindem2020control} \\
\bottomrule
\end{tabular}
\end{table*}

\end{document}